\documentclass[iop]{emulateapj}
\usepackage{ifthen}
\usepackage{natbib}

\newboolean{astroph}
\setboolean{astroph}{true}

\usepackage{color}
\newcommand{\ed}[1]{#1}
\newcommand{\rr}[1]{#1}

\newcommand{\Msun}{\ensuremath{\mathrm{M}_\odot}}

\newcommand{\um}{\ensuremath{\mathrm{\mu m}}}

\newcommand{\kmps}{\ensuremath{\mathrm{km\ s}^{-1}}\ }

\newcommand{\kpchone}{\ensuremath{h_{100}^{-1}~\mathrm{kpc}}}
\newcommand{\Gyr}{\ensuremath{\mathrm{Gyr}}}
\newcommand{\kpc}{\ensuremath{\mathrm{kpc}}}

\newcommand{\logMsun}{\ensuremath{\log M/\Msun}}

\newcommand{\Tobs}{\ensuremath{T_{\mathrm{obs}}}}
\newcommand{\Cmerge}{\ensuremath{C_{\mathrm{merge}}}}
\newcommand{\Clos}{\ensuremath{C_{\mathrm{l.o.s.}}}}
\newcommand{\fecp}{\ensuremath{f_{\mathrm{merge}}}}
\newcommand{\Rmerge}{\ensuremath{\Re_{\mathrm{merge}}}}

\newcommand{\figdemo}{
  \begin{figure*}[tbp]
    \centering
    \plotone{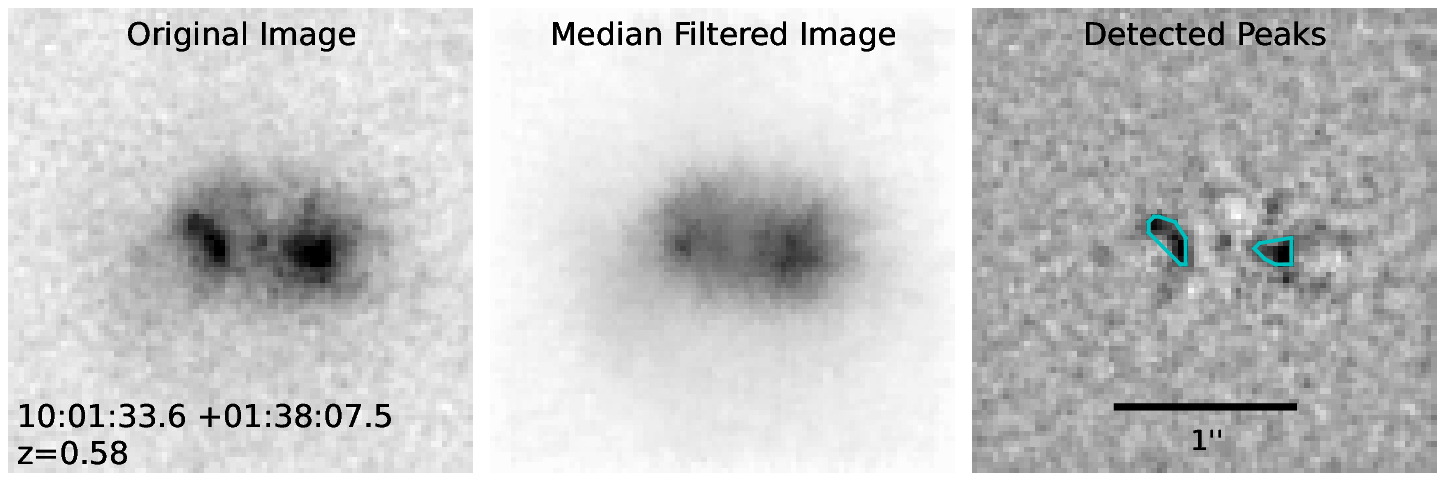}
    \caption
    {Demonstration of median filter and peak detection on an image of a merger. The (cyan) contours in the last panel outline the two detected peaks in the difference (original $-$ median-filtered) image. The peaks are separated by 4.0 \kpc{} and have a flux ratio of 1:1. }
    \label{fig:demo}
  \end{figure*}
}

\newcommand{\figexample}{
\begin{figure*}[tbp]
\centering
\plotone{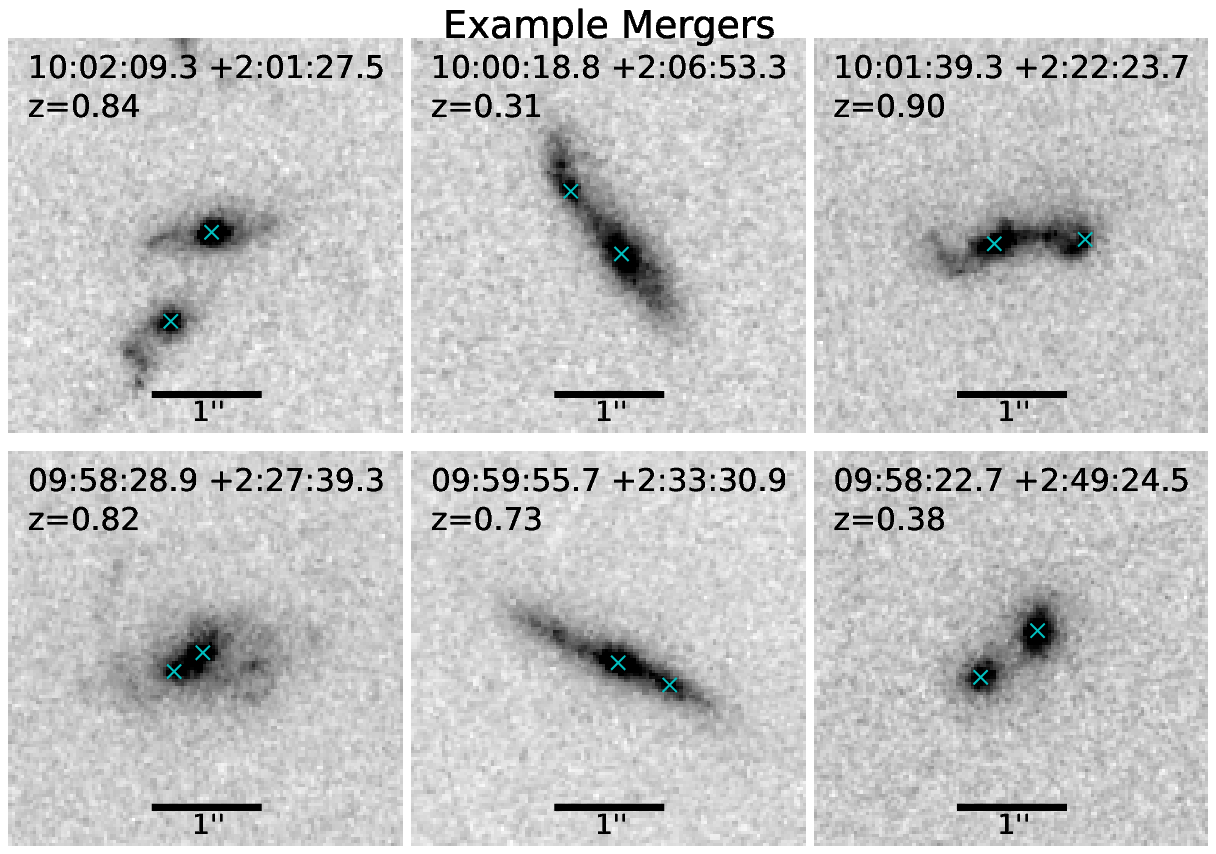}
\caption{Examples of merging galaxies in the photo-z sample after cuts in peak separation and peak flux. The $\times$s show peaks found by the median fing filter. The bottom center image \rr{may be an edge-on disk with asymmetric spiral arms instead of a merger.} 
}
\label{fig:example}
\end{figure*}
}

\newcommand{\figreject}{
\begin{figure*}[tbp]
\centering
\plotone{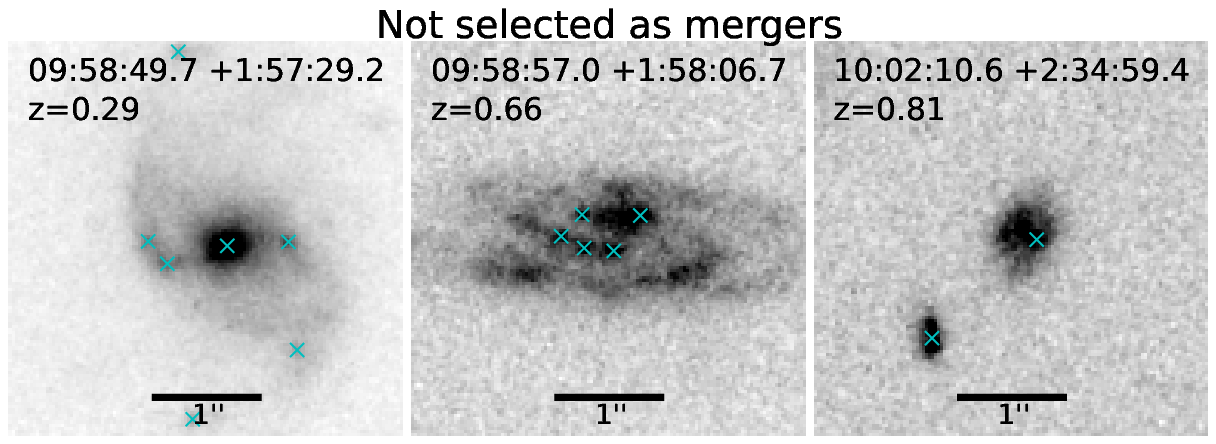}
\caption{Examples of galaxies with multiple peaks detected, but that are not considered mergers. The $\times$s show peaks found by the median ring filter. These galaxies are removed from the merger sample by the cuts explained in \S\ref{sec:dblnuc}. These galaxies fail because the detected peaks are too faint compared to the central peak (left), all but the central peak are too faint compared to the whole galaxy (middle), and the two peaks have a projected separation larger than $8\ \kpc$ (right).
}
\label{fig:reject}
\end{figure*}
}

\newcommand{\figginiM}{
  \begin{figure}[tbp]
    \centering \plotone{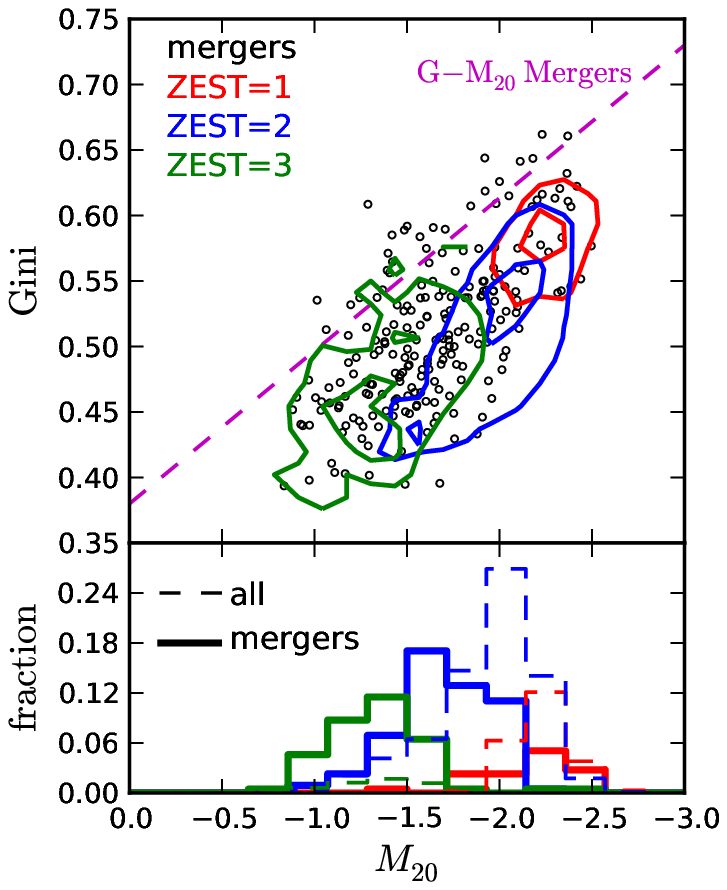}
    \caption
    {{\it Top}: Gini coefficient and $M_{20}$ values for late-stage mergers (black points). The contours show the parent sample color-coded by ZEST galaxy type \citep{Scarlata2007}. The sample is limited to galaxies with stellar masses greater than $2.5\times 10^{10}\, \Msun$. ZEST=1,2,3 are ellipticals, spirals (with bulges), and irregulars, respectively. The inner(outer) contours contain $30\%$($80\%$) of the galaxies of each ZEST type. The dashed magenta line is the criterion for merging galaxies from \citet{Lotz2004}. Most late-stage mergers lie below this line and would not be detected using the $G-M_{20}$ method. {\it Bottom:} Distribution of $M_{20}$ for our sample of mergers (solid lines) and the parent sample (dashed lines). Colors indicate ZEST type as in the top panel. }
    \label{fig:giniM20}
  \end{figure}
}
\newcommand{\figconcenasym}{
  \begin{figure}[tbp]
    \centering
    \plotone{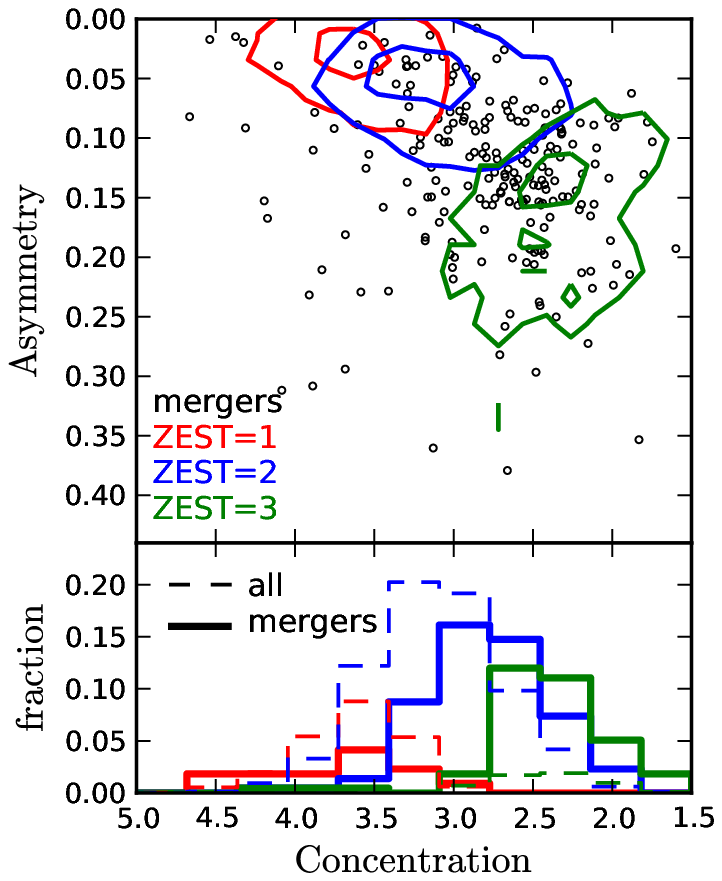}
    \caption
    {{\it Top:} Petrosian concentration and asymmetry (about $180\,^{\circ}$ rotation). Symbols and contours are as in Figure \ref{fig:giniM20}. Half of the late-stage mergers are of ZEST type $2$ (spirals). A late-stage merger is $7$ times more likely to be an irregular galaxy (ZEST=3) than a galaxy from the parent sample. {\it Bottom:} Distribution of concentration for mergers (solid lines) and the parent sample (dashed lines). Colors indicate ZEST type as in the top panel. On average, late-stage mergers of ZEST type 2 (blue lines) are less concentrated than typical ZEST=2 galaxies. } 
    \label{fig:concenasym}
  \end{figure}
}

\newcommand{\figmasscomp}{
\begin{figure}[btp]
  \centering
  \epsscale{1.25}
  \plotone{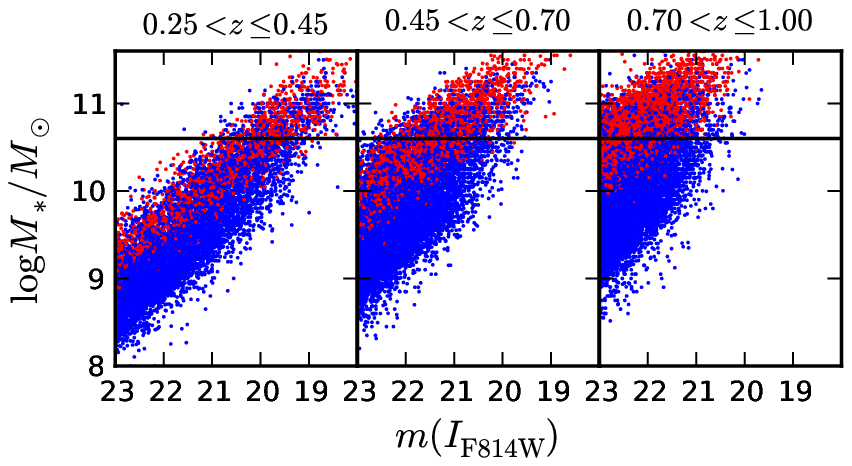}
  \caption{\rr{The mass completeness of our photometric sample ($I<23$) as a function of $I_{\mathrm{814W}}$ apparent magnitude in three redshift bins. Quescient(star-forming) (see \S\ref{ssec:mcolor}) galaxies are shown in red(blue). Up to $z=0.7$, the sample is complete for both populations. Beyond $z=0.7$, the completeness drops to $\sim 90\%$, with most of the missing galaxies above $z=0.9$.}
}
\label{fig:masscomp}
\end{figure}
}

\newcommand{\figmergerate}{
\begin{figure}[tbp]
  \centering
  \plotone{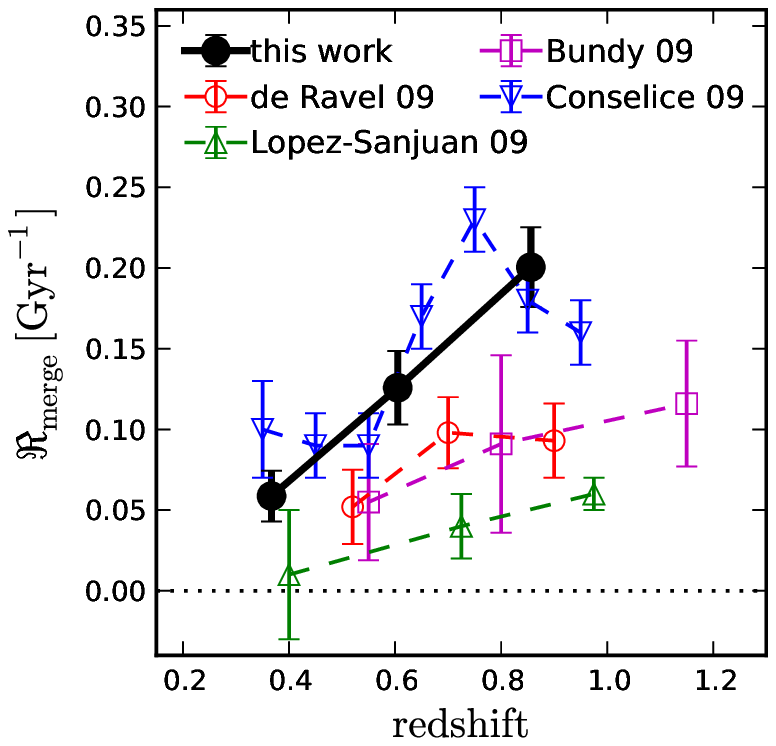}
  \caption{The fractional merger rate as a function of redshift derived from our sample of late-stage mergers (black, filled circles), compared to other stellar mass-selected studies of merging galaxies. \citet{Bundy2009} and \citet{deRavel2009} identify merging galaxies as photometric pairs. \citet{LopezSanjuan2009, Conselice2009} identify mergers based on galaxy asymmetry. All data is from \citet{Lotz2011} in which the timescales for observation of mergers are calibrated to facilitate easy comparison. Our merger rates are corrected from line-of-sight superpositions, contamination and incompleteness. The error bars on our points (black) are the statistical error bars only, but include the uncertainties in the correction for contamination and incompleteness.}
  \label{fig:mergerate}
\end{figure}
}

\newcommand{\fignuvrj}{
  \begin{figure}[tbp]
\epsscale{1.25}
\plotone{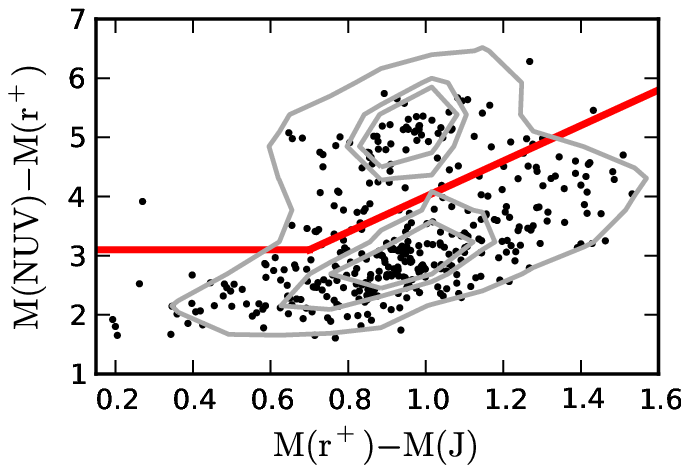}
\caption{Color-color diagram for selecting queiscent and star-forming galaxies. The cuts are from \citet{Ilbert2013, Ilbert2010}. Queiscent galaxies are in the upper left. The data shown is in the redshift range $0.25 < z < 1.0$ and with stellar masses $>10^{10}\Msun$. The contours show the distribution of the full photo-z sample, with contours at the 30$^{\mathrm{th}}$, 50$^{\mathrm{th}}$, and 90$^{\mathrm{th}}$ percentiles. The points show the late-stage mergers.
}
\label{fig:nuvrj}
\end{figure}
}

\newcommand{\figmergecolormass}{
\begin{figure*}[tbp]
\centering
\plotone{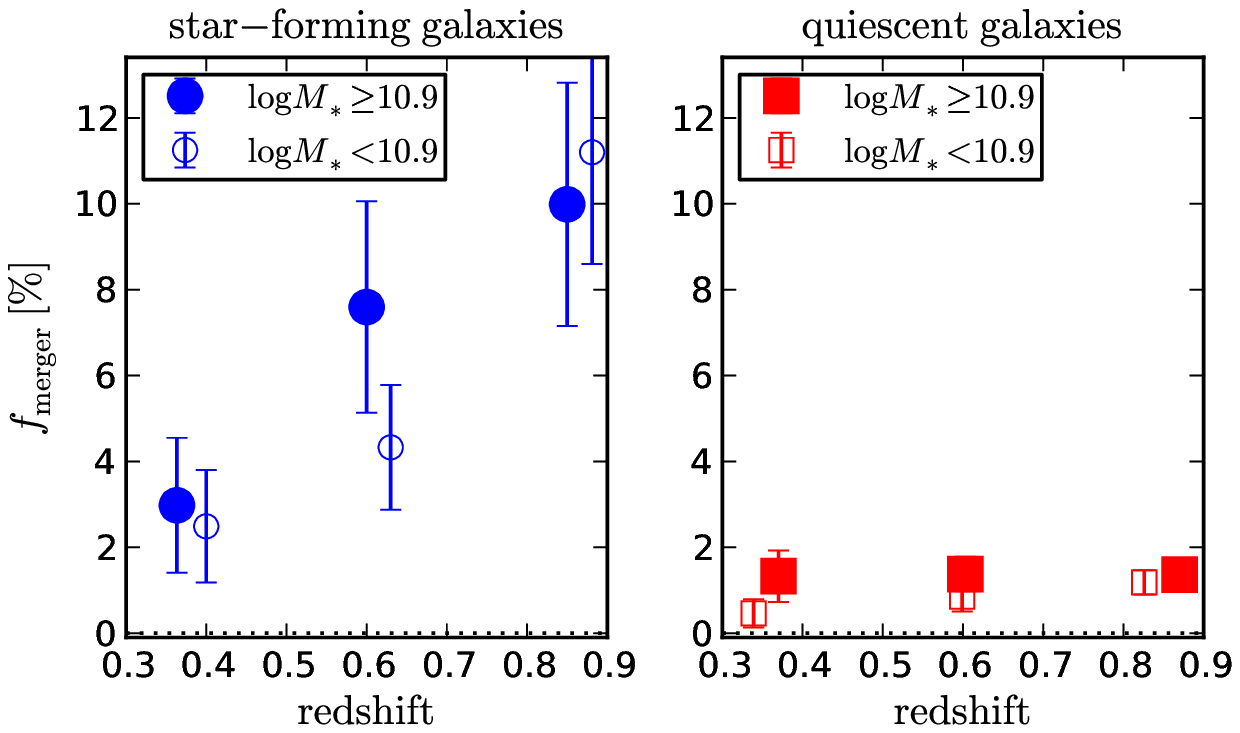}
\caption{The fraction of mergers for different types of galaxies,\rr{ corrected for chance line-of-sight superpositions, contamination, and incompleteness}, as a function of redshift. The left panel shows star-forming galaxies while the right panel shows queiscent galaxies, separated by the color cuts in Figure \ref{fig:nuvrj}. The solid (empty) symbols show high (low) mass galaxies. Small horizontal offsets are added for visibility.
}
\label{fig:mergecolormass}
\end{figure*}
}

\newcommand{\figmergecolorfrac}{
\begin{figure}[tbp]
\centering
\plotone{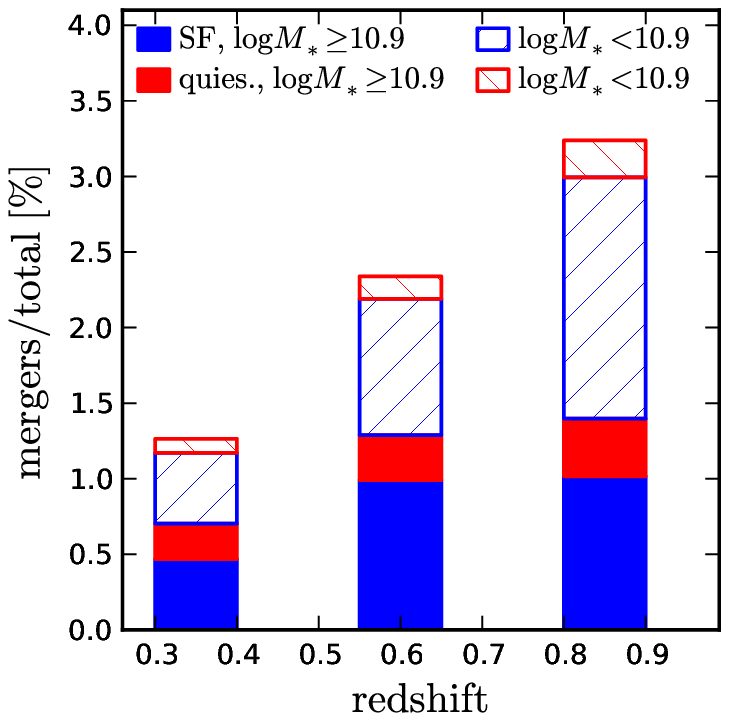}
\caption{\rr{The fraction of pairs split by galaxy color and mass, corrected for line-of-sight superpositions, incompleteness and contamination. With these corrections, mergers between star-forming galaxies dominant at all redshifts, and the number of star-forming mergers increases significantly with redshift. }
}
\label{fig:mergecolorfrac}
\end{figure}
}

\newcommand{\figmergecolorfracnc}{
\begin{figure}[tbp]
\centering
\plotone{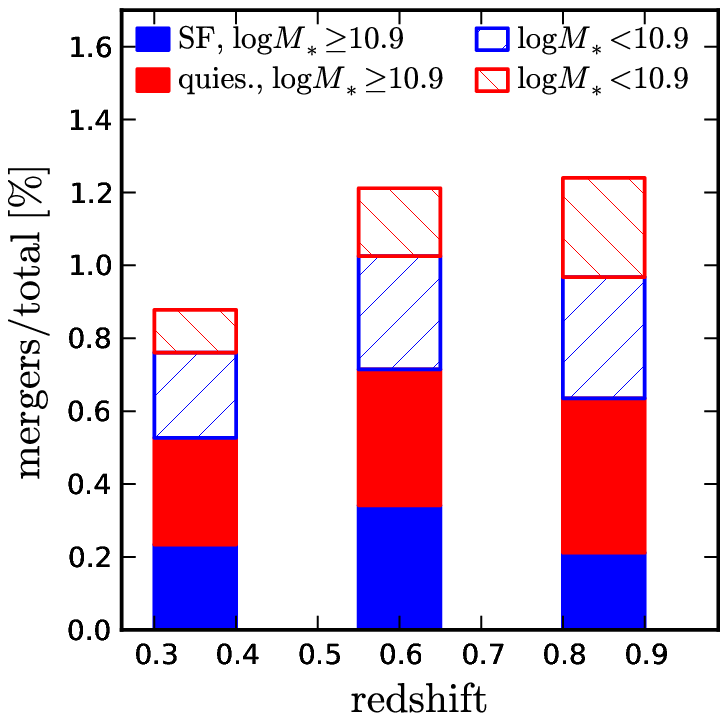}
\caption{\rr{As in Figure \ref{fig:mergecolorfrac}, but the fractions are \emph{not} corrected for incompleteness or contamination.  Without the incompleteness corrections, there is no statistically significant increase in merger activity at high redshifts. Because the correction factor is small for early-type galaxies, the fraction of queiscent mergers is nearly the same as in Figure \ref{fig:mergecolorfrac}.}
}
\label{fig:mergecolorfracnc}
\end{figure}
}

\newcommand{\figsSFR}{
  \begin{figure}[tbp]
    \centering
    \plotone{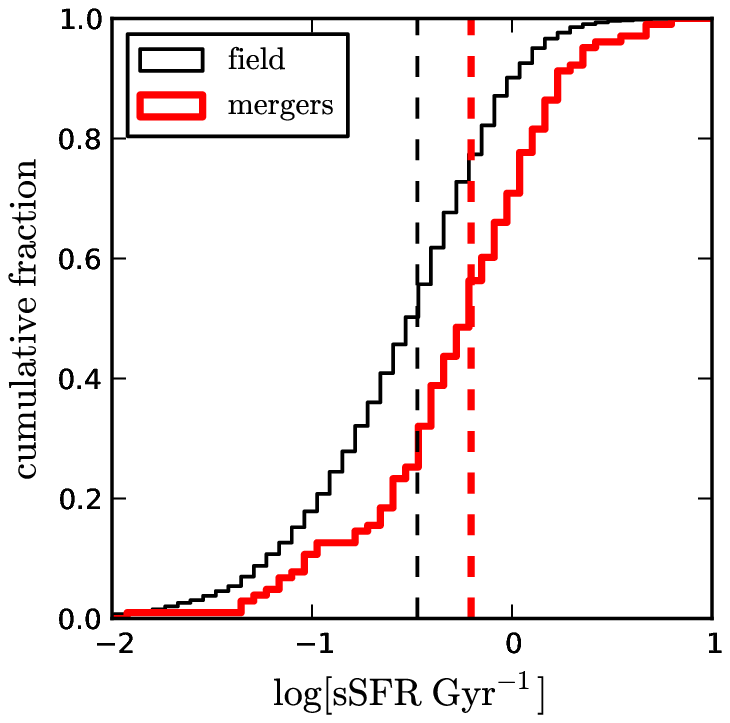}
    \caption
        {The cumulative distribution of sSFR for late-stage mergers and non-interacting galaxies from the zCOSMOS sample. \rr{The sSFR is derived from the {\it Spitzer}/MIPS $24\um$ flux. The median sSFR in late-stage mergers is a factor of $2.1\pm0.6$ larger than that in non-interacting galaxies.}
}
   \label{fig:cumsSFR}
  \end{figure}
}

\newcommand{\figmassSFR}{
  \begin{figure}[tbp]
    \centering
    \plotone{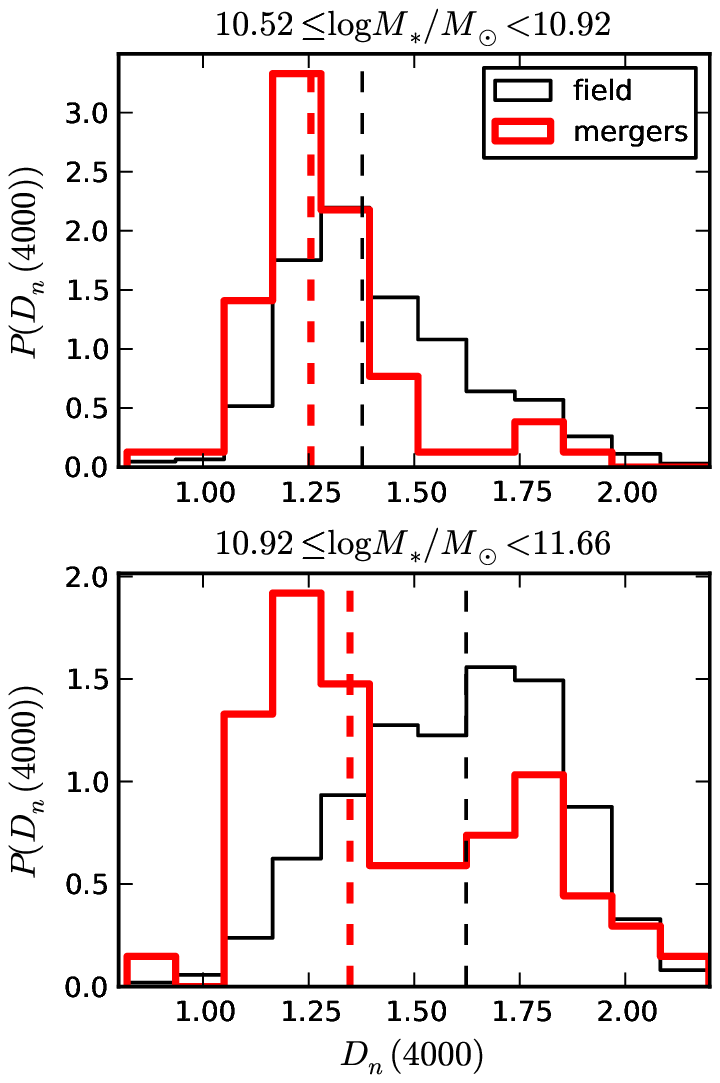}
    \caption{The distribution of the narrow $4000$~\AA-break for late-stage mergers (red) and the parent sample (black) from zCOSMOS. The galaxies are divided into two mass bins. In the upper(lower) panel there are \rr{1582(2375)} control galaxies and \rr{69(67)} merging galaxies. The corresponding vertical lines show the medians of each distribution. The median $D_n(4000)$ for late-stage mergers is smaller in both mass bins, indicating most late-stage mergers have undergone recent star formation. }
        \label{fig:massSFR}
  \end{figure}
}

\newcommand{\figAGNex}{
\begin{figure*}[tpb]
\centering
\plotone{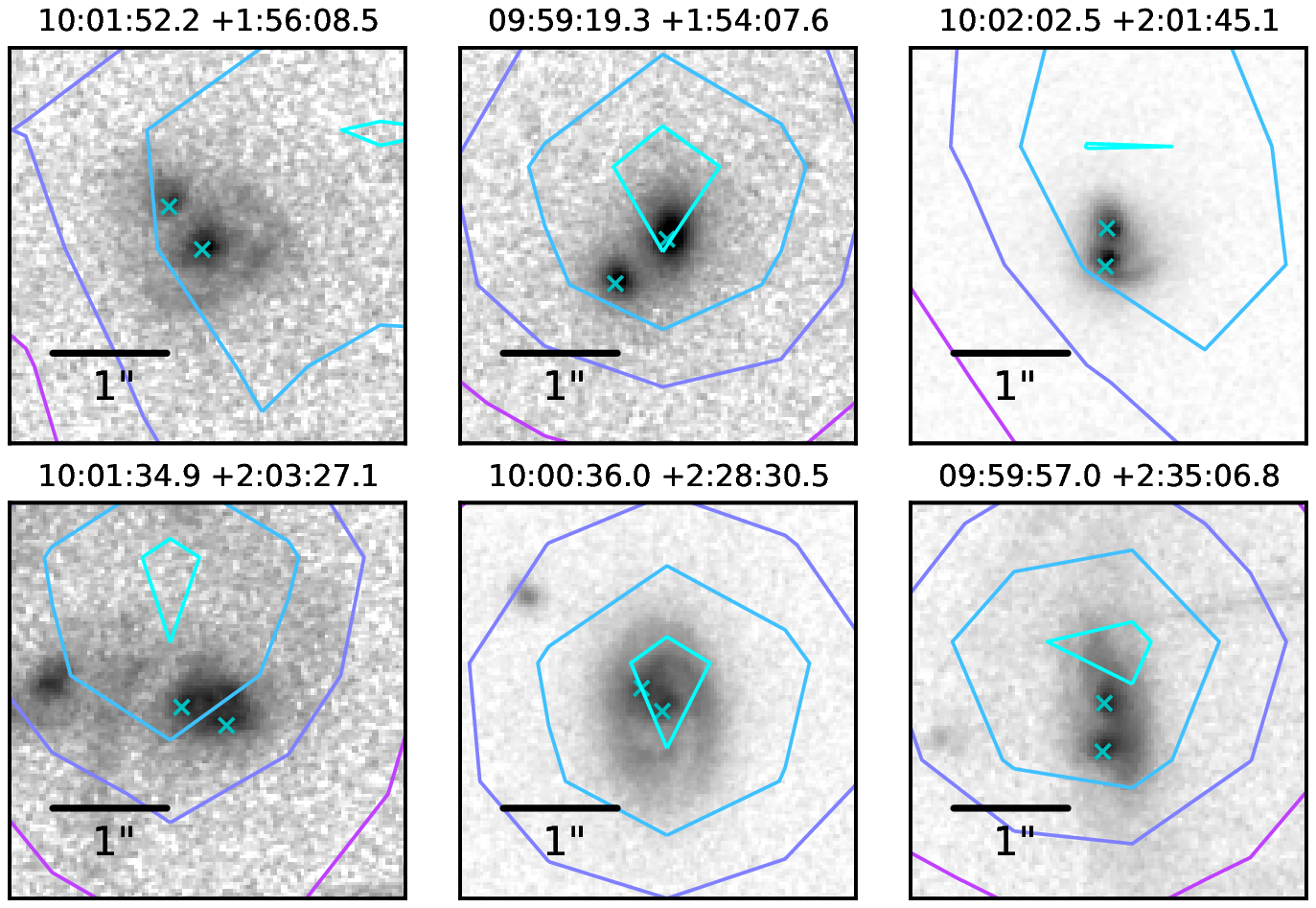}
\caption{Images of the six late-stage mergers that are also X-ray selected AGN. The (blue) contours show the total ($0.5-7.0$~keV) flux from {\it Chandra} \citep{Elvis2009}.  The (cyan) crosses show the position of the two peaks found by our merger-finding method. The galaxy in the lower middle panel may be a spiral galaxy. }
\label{fig:agnex}
\end{figure*}
}

\newcommand{\figAGNfrac}{
\begin{figure}[tbp]
\includegraphics[scale=1.05, clip=true]{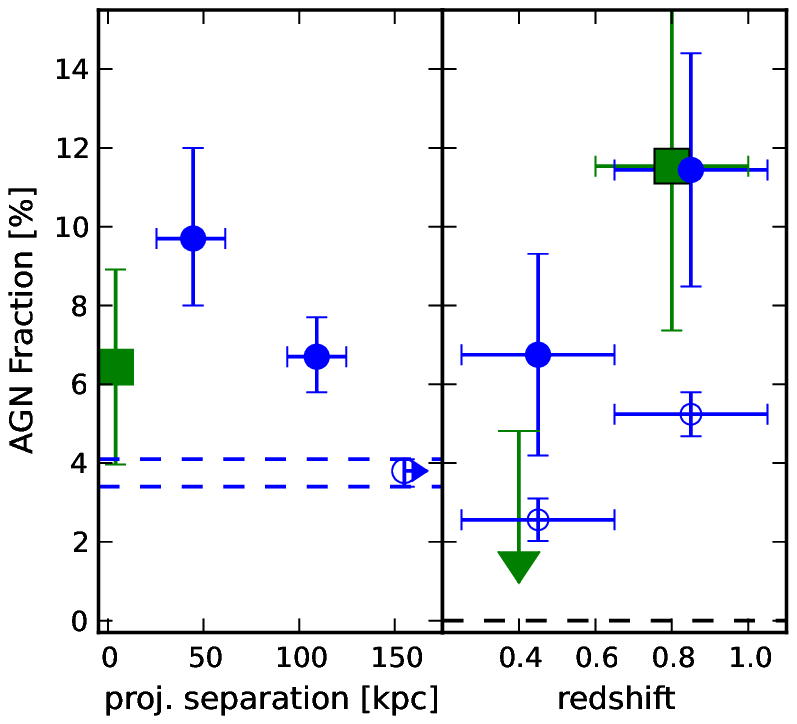}
\caption{{\it Left:} The fraction of AGN in galaxy pairs at various projected separations. The $3$ right points are from \citet{Silverman2011} \ed{(see their Figure 5)}. The empty symbol is the field value, corrected for unidentified kinematic pairs. The square shows the AGN fraction in late-stage mergers ($6$ mergers). \ed{The points are plotted at the median separation in each bin, and the horizontal error bars denote the interquartile range ($25\%-75\%$) of the galaxies in each bin.} {\it Right:} The AGN fraction in pairs in two redshift bins. The squares denote late-stage mergers. There are no late-stage mergers in the low redshift bin and the error bar denotes the $1\sigma$ upper limit. The filled and empty circles are the AGN fraction in pairs separated by $<75 \kpc$, and the field, respectively, and are taken from \citet{Silverman2011}. Note that in this panel, the field AGN fractions are not corrected for contamination by kinematic pairs and should be $\sim 0.5-1\%$ lower. }
    \label{fig:agnfrac}
  \end{figure}
}

\newcommand{\figmocks}{
\begin{figure}[hbtp]
\centering
\includegraphics{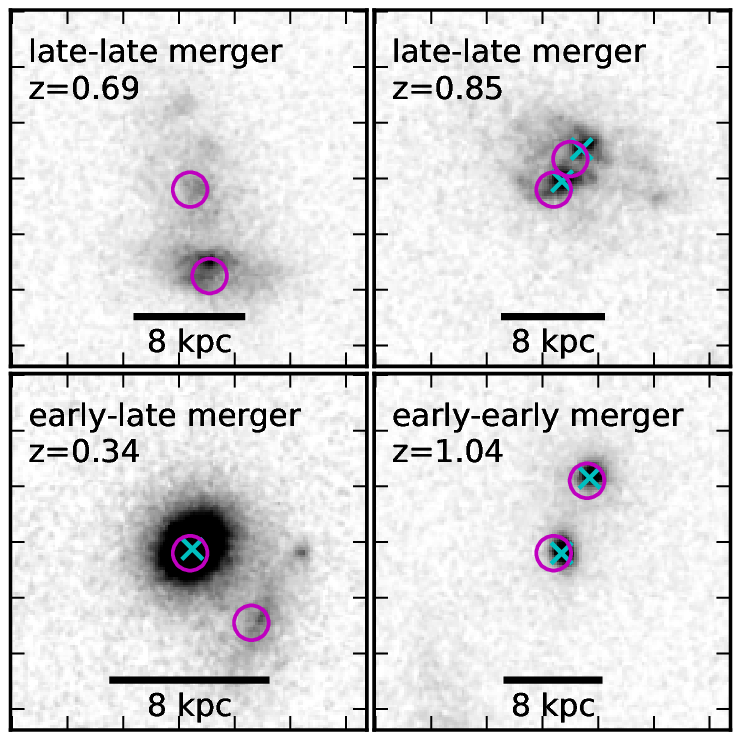}
\caption{Example mock merger images. The circles show the position of the coadded galaxies, while the crosses show the positions of the detected peaks. Both galaxy mergers on the right are sucessfully detected and pass the cuts implemented to remove contaminants. For the merger in the upper left, both galaxies are late-type and too diffuse to be detected. For the merger in the lower left, the flux ratio between the two galaxies is too large to be detected.
}
\label{fig:mocks}
\end{figure}
}

\newcommand{\figzcomplete}{
\begin{figure}[htpb]
\centering
\plottwo{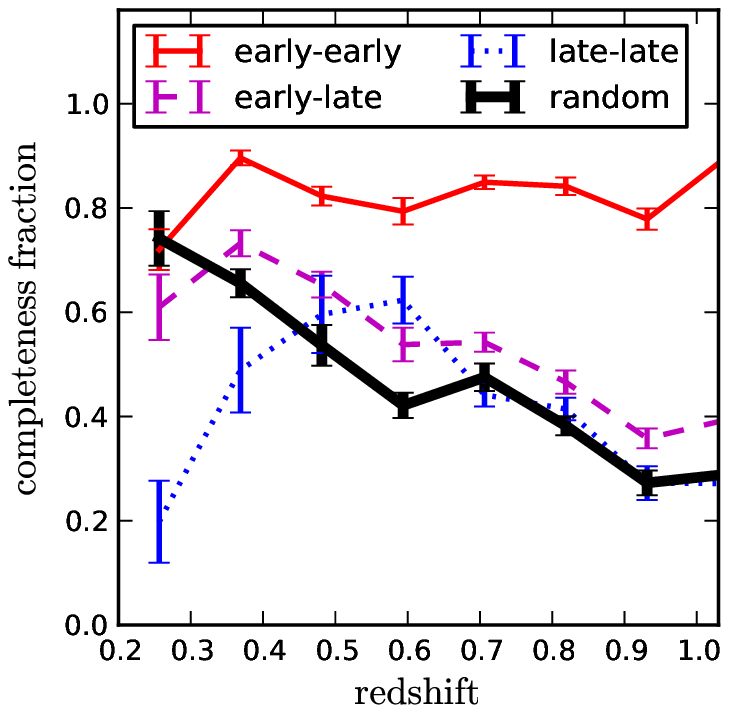}{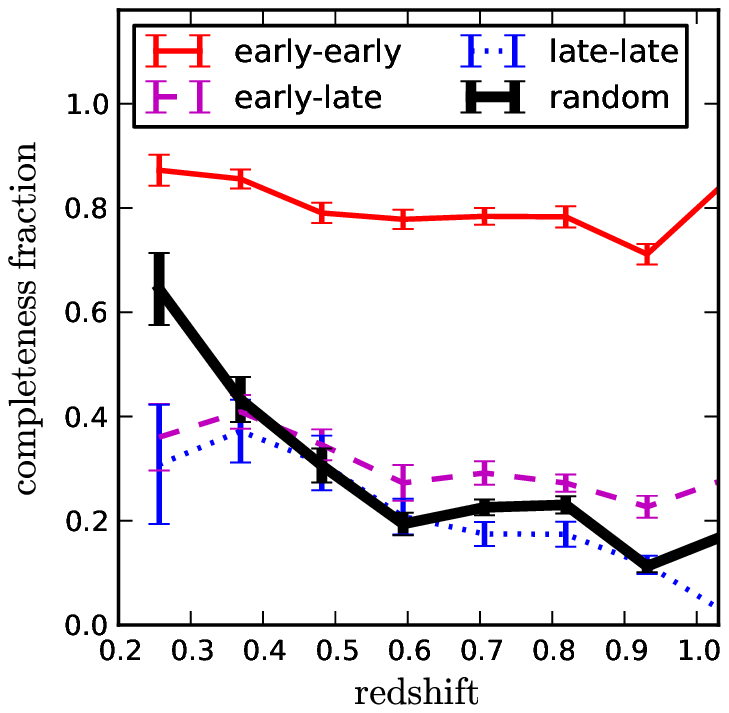}
\caption{The completeness of the late-stage major mergers in simulated images as a function of redshift, before applying cuts in flux ratio for contamination (left) and after applying cuts (right). The thick black line shows the completeness for a random sample of galaxies, with a representative morphological mix. While the other lines show the completeness for mock mergers in which the merging galaxies are both early types (red, solid), both late types (blue, dotted), and mixed (magenta, dashed). The morphologies are determined by the ZEST parameter. The error are derived by bootstrap resampling. The completeness is a stronger function of morphology than redshift.
}
\label{fig:zcomplete}
\end{figure}
}

\newcommand{\figscomplete}{
\begin{figure}[hbtp]
\centering
\includegraphics{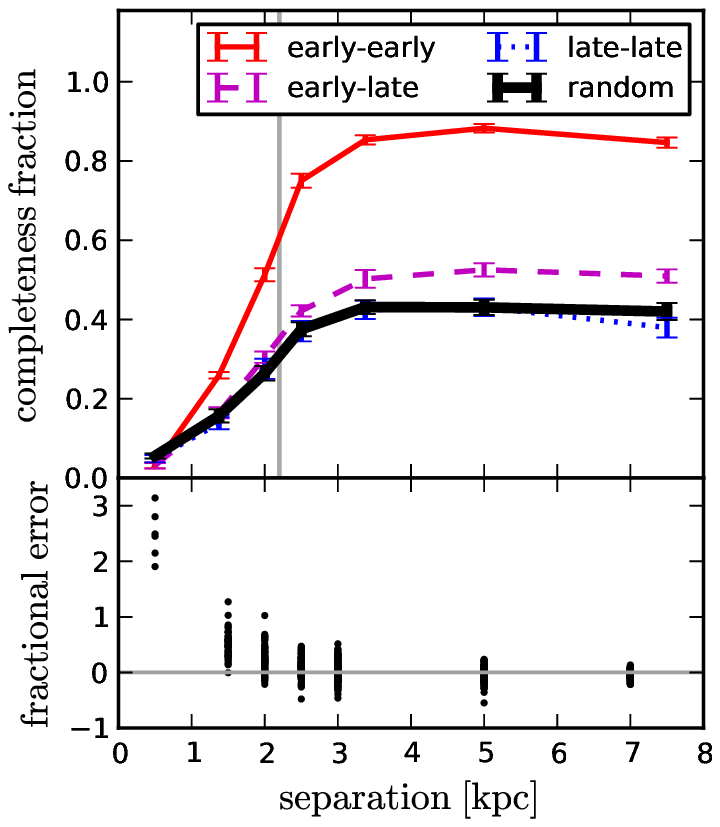}
\caption{{\it Top}: The completeness of the mock mergers as a function of pair separation. Note that each galaxy pair was simulated at a discreet set of separations. The completeness drops sharply for separations comparable to the median ring filter size. The different lines show the completeness for different morphologies for the members of the merger, as in Figure \ref{fig:zcomplete}.  The vertical line shows the cut made in separation at $2.2\ \kpc$. {\it Bottom:} The fractional error in the measured peak separation compared to the real separation for the `random' sample of morphologies only. The separation is reasonably well measured beyond a few kpc. However, small separations are typically overestimated which will lead to contamination of our sample by mergers with separations $ < 2.2$  \kpc{}.
}
\label{fig:scomplete}
\end{figure}
}

\newcommand{\figfluxfrac}{
\begin{figure}[hbtp]
\centering
\includegraphics{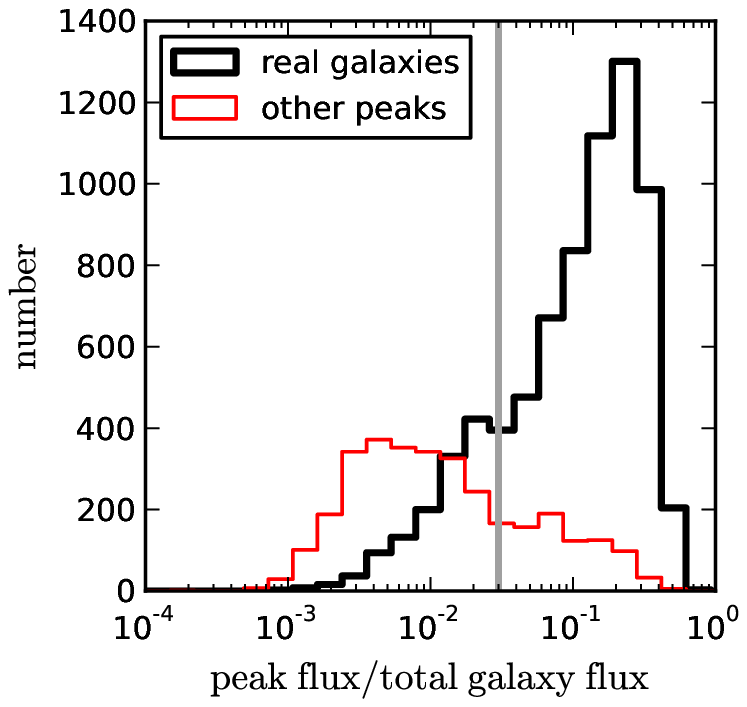}
\caption{Distribution of the detected peak flux ratios. The thick line histogram shows the distribution of peak to total flux for detected galaxy sources in the mock merger images. The thin line shows the distribution of extraneous peaks detected. By putting a cutoff at $3\%$, the contamination from extraneous peaks is $10\%$.
}
\label{fig:fluxfrac}
\end{figure}
}

\newcommand{\figfcontaim}{
\begin{figure}[hbtp]
\centering
\includegraphics{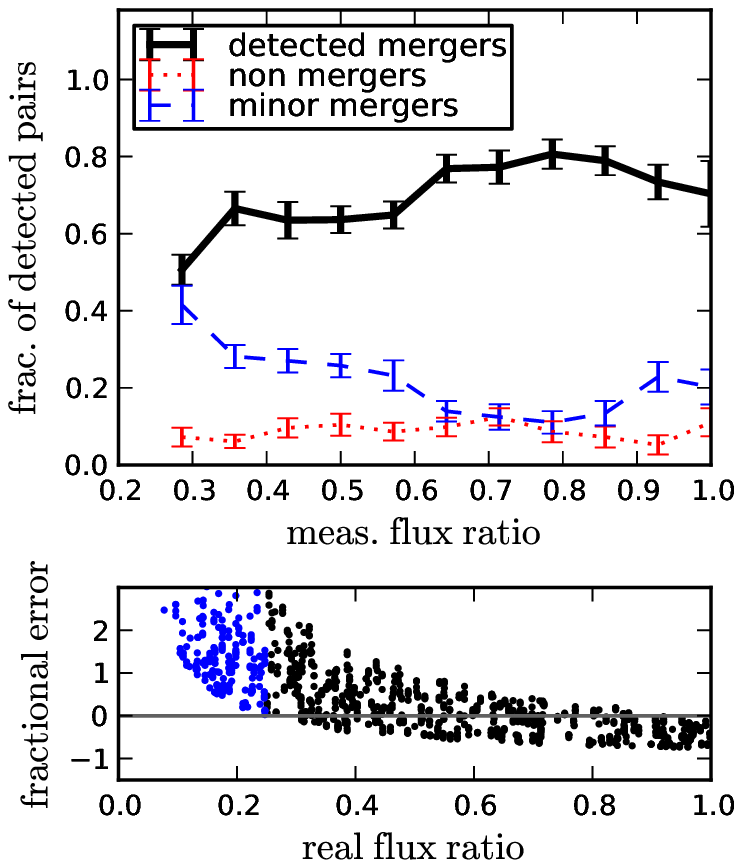}
\caption{{\it Top:} The fraction of detected mergers which are major mergers (solid line), minor mergers (dashed line) and contaminants (dotted lines). The major (minor) mergers consist of two galaxies with a flux ratio larger (smaller) than $0.25$. The contaminants are detected late-stage mergers which do not match the mock galaxies placed in each image. These include detections of galactic substructure. {\it Bottom:} The fractional error in the measured flux ratio as a function of real flux ratio. For small real flux ratios, our method typically overestimates small flux ratios, which leads to a contaimination from minor mergers. 
}
\label{fig:fcontaim}
\end{figure}
}

\newcommand{\fignotgini}{
\begin{figure}[htpb]
\centering
\plotone{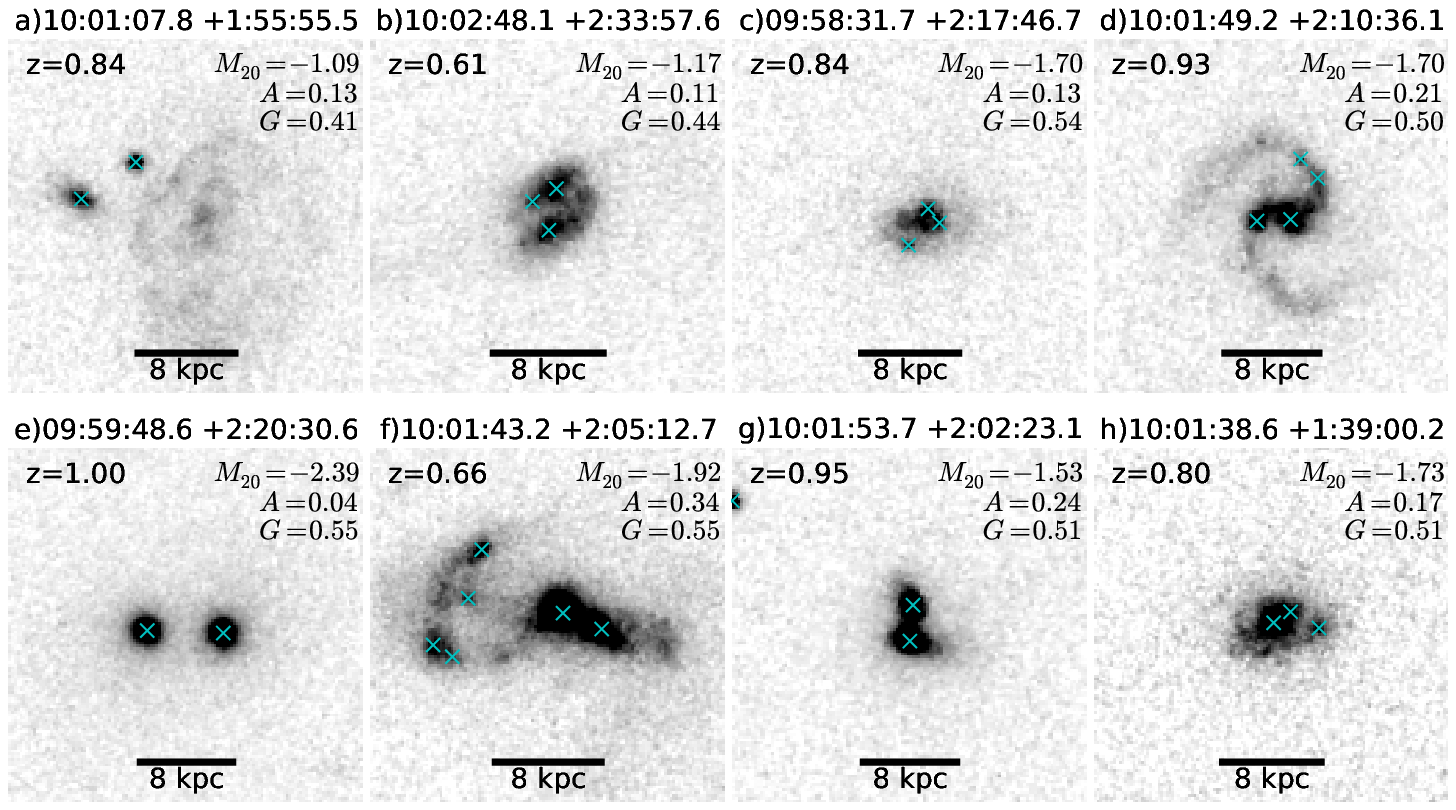}
\caption{Examples of galaxies which are late-stage mergers but are \emph{not} detected as mergers by the Gini-$M_{20}$ method \citep[][]{Lotz2008}.  We only examine galaxies with $z>0.6$ in order to minimize the effects of morphological k-corrections to the rest frame $B-$band. Crosses show all detected peaks, before any cuts on projected separation or flux ratio. The images are shown with an arcsinh stretch and with the same scaling.
}
\label{fig:notgini}
\end{figure}
}

\newcommand{\fignotasym}{
\begin{figure}[htpb]
\centering
\plotone{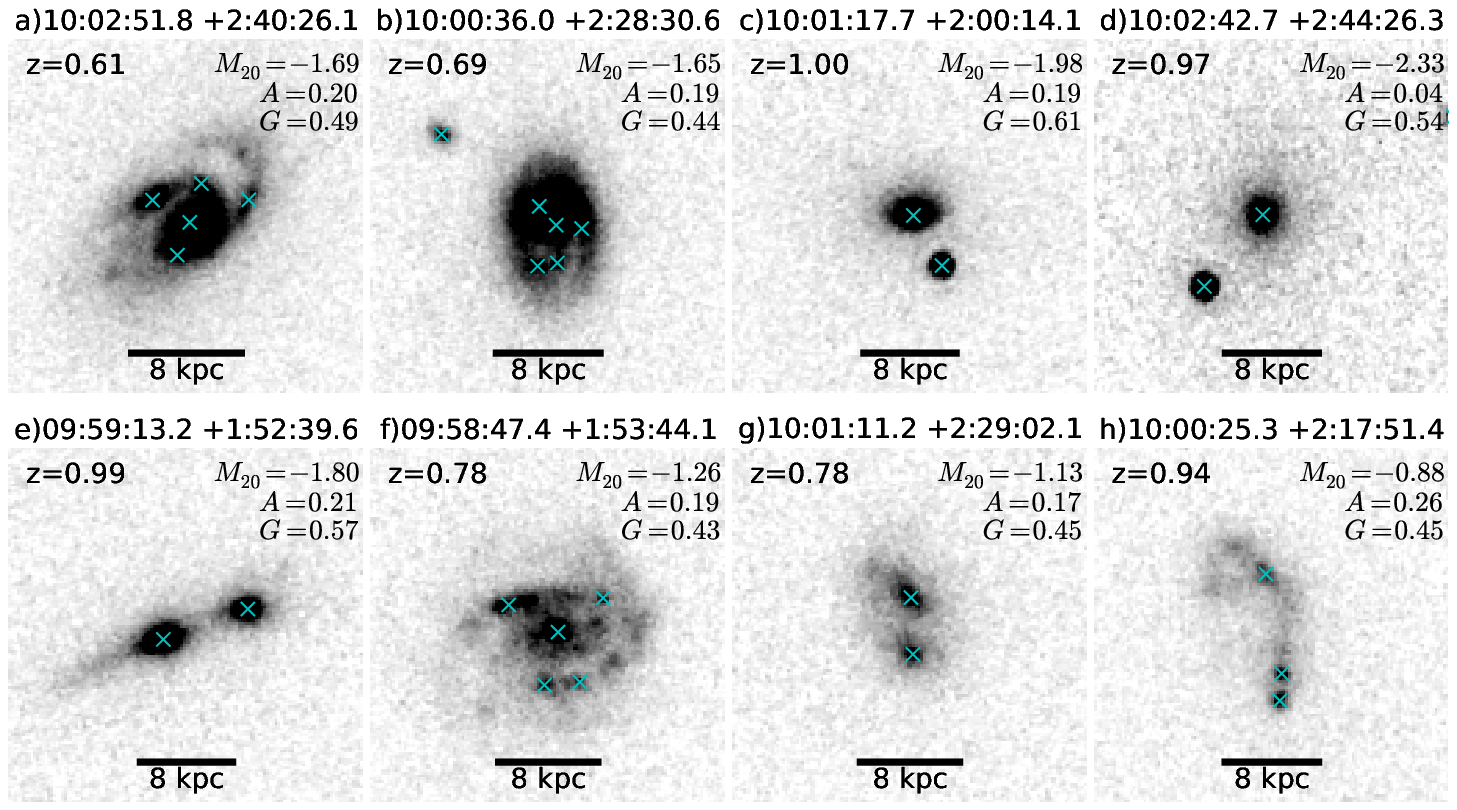}
\caption{Examples of galaxies which are late-stage mergers but are \emph{not} selected as mergers based on their asymmetry around a $180^\circ$ rotation ($A>0.35$, \citep[see][]{Conselice2003}). We use the asymmetry measurements from \citet{Cassata2005} and include a correction of $0.05$ for the surface brightness dimming \citep{Conselice2009}. We only use galaxies with $z>0.6$, which limits the morphological k-corrections when comparing to the rest frame $B$-band. The images have the same stretch and scaling as those in Figure \ref{fig:notgini}.
}
\label{fig:notasym}
\end{figure}
}

\newcommand{\figisgini}{
\begin{figure}[htpb]
\centering
\plotone{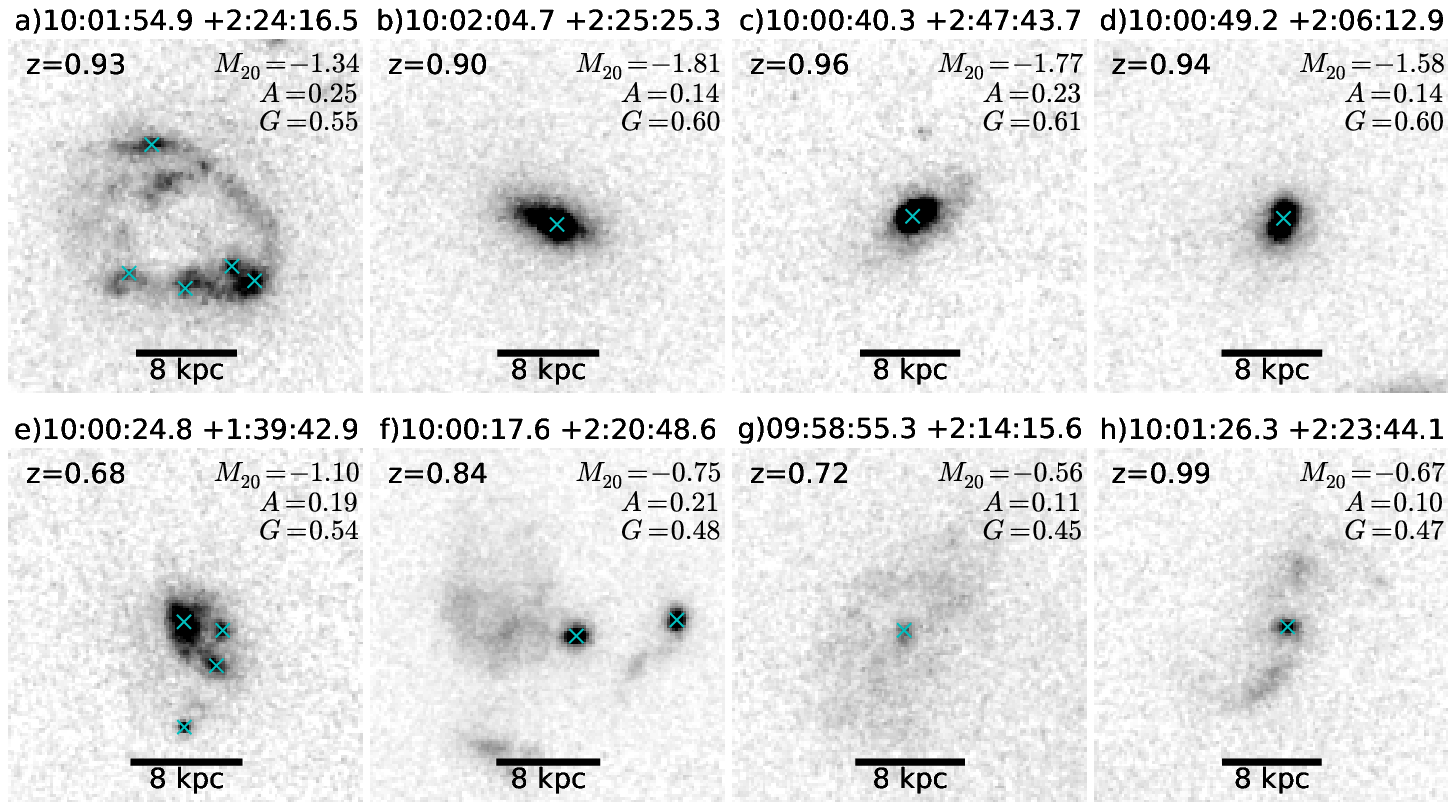}
\caption{Examples of galaxies which are \emph{not} late-stage mergers but are selected as mergers by the Gini-$M_{20}$ method \citep[][]{Lotz2008}. The redshift range is the same as in Figure \ref{fig:notgini} The images have the same stretch and scaling as those in Figure \ref{fig:notgini}. Most of these systems would be characterized as minor mergers, and therefore missed by our method.
}
\label{fig:isgini}
\end{figure}
}

\newcommand{\figisasym}{
\begin{figure}[htpb]
\centering
\plotone{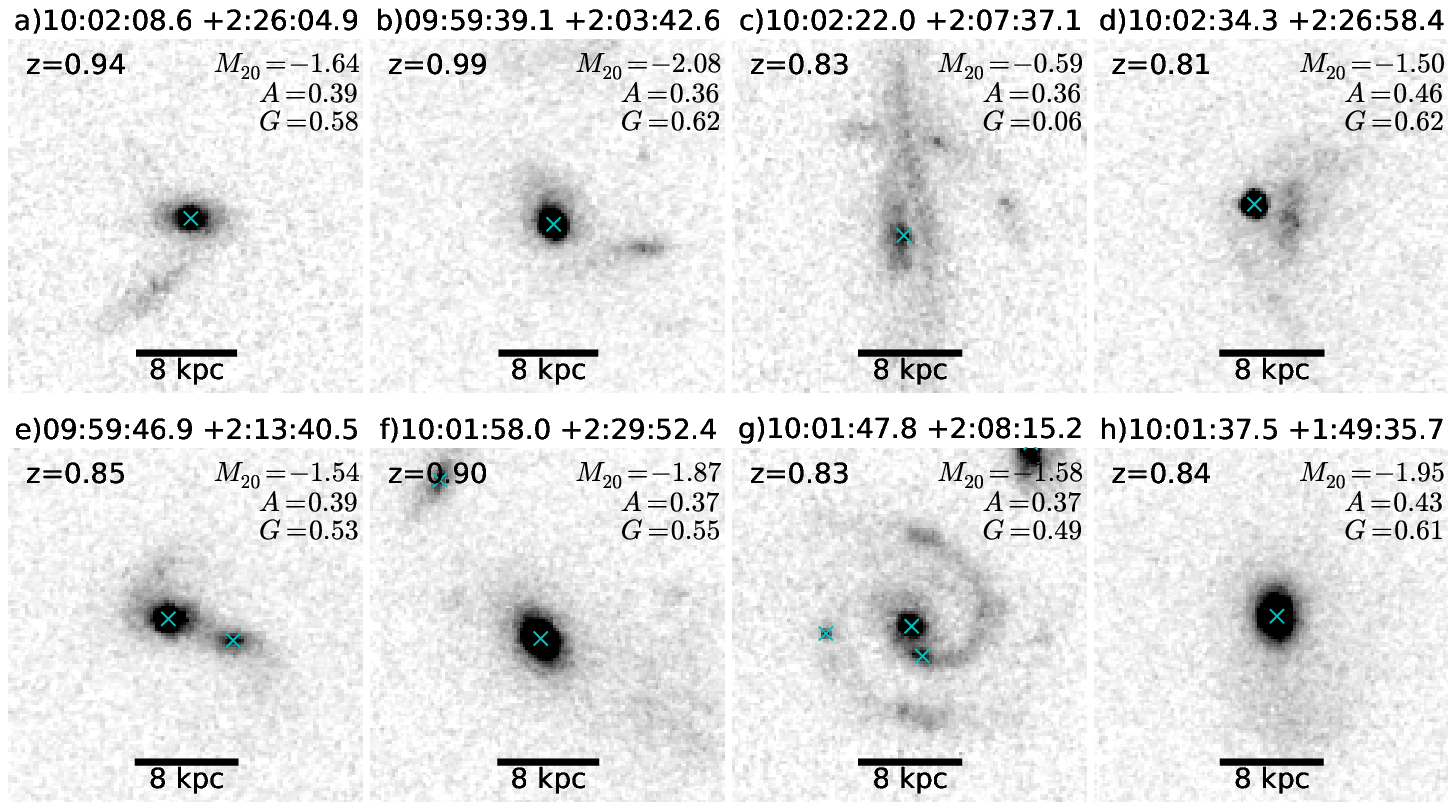}
\caption{Examples of galaxies which are \emph{not} late-stage mergers but are selected as mergers based on their asymmetry around a $180^\circ$ rotation ($A>0.35$ \citep[see][]{Conselice2003}). The images have the same stretch and scaling as those in Figure \ref{fig:notgini}.}
\label{fig:isasym}
\end{figure}
}

\newcommand{\tablesamps}{
\begin{deluxetable*}{lccccccl}
\tablecaption{Parent sample properties\label{tabsamps}}
\tablehead{\colhead{Parent sample} & \colhead{$m(I_{\mathrm{814W}})$ limit} & \colhead{$z$ limits} & \colhead{$\log M_*/M_\odot$ limit} & \colhead{N$_{\mathrm{gal}}$} & \colhead{N$_{\mathrm{merger}}$} & \colhead{pair~sep.~limits~[kpc]} & \colhead{section}}
\startdata
photo-z & $<23$ & -- & -- & 44164 & 2047 & $<8$ & \S\ref{ssec:photosample}, Table \ref{tabp}\tablenotemark{a} \\
photo-z & $<23$ & -- & -- & 44164 & 1547 & [2.2,8] & \S\ref{ssec:photosample} \\
photo-z & $<23$ & [0.25,1.0] & $>10.6$ & 6226 & 148 & [2.2,8] & \S\ref{sec:mergerrates} \\
spec-z & $<22.5$ & [0.25, 1.05] & $>10.4$ & 5001 & 166 & $<8$ & \S\ref{ssec:specsample}, \ref{ssec:sfr} \\
spec-z & $<22.5$ & [0.25, 1.05] & $>10.4$ & 3474 & 112 & $<8$ & \S\ref{ssec:specsample}, \ref{ssec:agn}\tablenotemark{b} \\
\enddata
\tablenotetext{a}{The full sample without any cuts in redshift, mass, or pair separation is very incomplete at high redshift. We do not use it for any analysis.}
\tablenotetext{b}{The smaller spectroscopic sample overlaps with the {\it Chandra} survey \citep{Elvis2009}, which is used to select X-ray AGN.}
\end{deluxetable*}
}

\newcommand{\tablepairs}{
\begin{deluxetable*}{llcccccccc}
\setlength{\tabcolsep}{0in}
\tabletypesize{\small}
\tablecaption{Late-stage mergers in photo-z + spec-z samples\label{tabp}}
\tablehead{\colhead{RA (J2000)\tablenotemark{a}} & \colhead{Dec (J2000)\tablenotemark{a}} & \colhead{photo-z\tablenotemark{b}} & \colhead{spec-z\tablenotemark{c}} & \colhead{$m_I$\tablenotemark{d}} & \colhead{$\log M_*/M_\odot$ \tablenotemark{e}} & 
\colhead{separation [\arcsec]} & \colhead{flux ratio} & \colhead{{\it Chand.} $\log L_{X}$\tablenotemark{f}} & \colhead{{\it XMM} $\log L_{X}$\tablenotemark{f}}}
\startdata
$149.51058$ & $2.74338$ & $0.49$ & -- & $22.21$ &  9.57 &  $0.35$ & $0.50$ & -- & -- \\
$149.83214$ & $1.94120$ & $0.66$ & 0.70 & $22.48$ &  9.50 &  $0.78$ & $0.61$ & -- & -- \\
$149.83058$ & $1.90214$ & $0.73$ & 0.73 & $21.25$ &  10.95 &  $0.62$ & $0.31$ & 42.6 & -- \\
$150.20693$ & $1.68028$ & $1.11$ & -- & $22.76$ &  10.80 &  $0.31$ & $0.88$ & 43.8 & 43.9 \\
$150.39549$ & $2.05754$ & -- & $0.96$ & $21.71$ &  10.57 &  $0.44$ & $0.96$ & 43.1 & -- \\
$150.51472$ & $2.59320$ & $0.37$ & -- & $22.71$ &  8.81 &  $0.36$ & $0.34$ & -- & -- \\
\enddata
\tablecomments{Table \ref{tabp} is published in its entirety in the electronic edition. A portion is shown here for guidance regarding its form and content.}
\tablenotetext{a}{From \citet{Ilbert2013}.}
\tablenotetext{b}{From \citet{Ilbert2013}, except for x-ray sources, which are from \citet{Salvato2011}.}
\tablenotetext{c}{spectroscopic redshift from zCOSMOS \citep{Lilly2007, Lilly2009}}
\tablenotetext{d}{{\it HST}/ACS FW814 AB magnitude from \citet{Leauthaud2007}.}
\tablenotetext{e}{Stellar masses for {\it XMM} sources from \citet{Bongiorno2012}; for sources \emph{without} a photo-z, from \citet{Bolzonella2010} and \citet{Pozzetti2010}; otherwise, from \citet{Ilbert2013}.}
\tablenotetext{f}{$\log L_X$ is the X-ray luminosity in the band $0.5-10$~keV in units of $\mathrm{erg\ s^{-1}}$. {\it XMM} data from \citet{Brusa2010}, {\it Chandra} data from \citet{Civano2012}.}
\end{deluxetable*}
}

\newcommand{\tablemr}{
\begin{deluxetable}{lrrccc}
\tabletypesize{\small}
\setlength{\tabcolsep}{0.0in} 
\tablecaption{Pair fractions and merger rates\label{tabMR}}
\tablehead{\colhead{$z$} & \colhead{$N_{\mathrm{gal}}$} & \colhead{$N_{\mathrm{mgr}}$} & \colhead{\Cmerge} & \colhead{$f_{\mathrm{merge}}$} & \colhead{$\Re\, [\Gyr^{-1}]$}}
\startdata
$\left[0.25,\ 0.45\right)$ & $867$ & $15$ & $1.1 \pm 0.1$ & $1.9 \pm 0.5\%$ & $5.9 \pm 1.6\%$ \\
$\left[0.45,\ 0.70\right)$ & $1644$ & $39$ & $1.8 \pm 0.1$ & $4.2 \pm 0.8\%$ & $12.6 \pm 2.3\%$ \\
$\left[0.70,\ 1.00\right)$ & $3383$ & $82$ & $2.7 \pm 0.1$ & $6.6 \pm 0.8\%$ & $20.1 \pm 2.5\%$ \\
\cutinhead{low mass ($10.6 < \logMsun < 10.9$) sample}
$\left[0.25,\ 0.45\right)$ & $399$ & $9$ & $1.1 \pm 0.1$ & $2.5 \pm 0.9\%$ & $7.7 \pm 2.6\%$ \\
$\left[0.45,\ 0.70\right)$ & $782$ & $23$ & $1.8 \pm 0.1$ & $5.1 \pm 1.2\%$ & $15.6 \pm 3.5\%$ \\
$\left[0.70,\ 1.00\right)$ & $1766$ & $42$ & $2.7 \pm 0.1$ & $6.5 \pm 1.1\%$ & $19.7 \pm 3.2\%$ \\
\cutinhead{high mass ($\logMsun \geq 10.9$) sample}
$\left[0.25,\ 0.45\right)$ & $468$ & $6$ & $1.1 \pm 0.1$ & $1.4 \pm 0.6\%$ & $4.4 \pm 1.8\%$ \\
$\left[0.45,\ 0.70\right)$ & $862$ & $16$ & $1.8 \pm 0.1$ & $3.2 \pm 0.9\%$ & $9.8 \pm 2.6\%$ \\
$\left[0.70,\ 1.00\right)$ & $1617$ & $40$ & $2.7 \pm 0.1$ & $6.8 \pm 1.1\%$ & $20.5 \pm 3.4\%$ \\
\cutinhead{star-forming sample}
$\left[0.25,\ 0.45\right)$ & $413$ & $8$ & $1.4 \pm 0.2$ & $2.7 \pm 1.1\%$ & $8.2 \pm 3.2\%$ \\
$\left[0.45,\ 0.70\right)$ & $763$ & $21$ & $2.0 \pm 0.2$ & $5.6 \pm 1.4\%$ & $16.9 \pm 4.1\%$ \\
$\left[0.70,\ 1.00\right)$ & $1131$ & $36$ & $3.4 \pm 0.3$ & $10.7 \pm 2.0\%$ & $32.4 \pm 6.1\%$ \\
\cutinhead{quiescent sample}
$\left[0.25,\ 0.45\right)$ & $454$ & $7$ & $0.6 \pm 0.0$ & $0.9 \pm 0.3\%$ & $2.6 \pm 1.0\%$ \\
$\left[0.45,\ 0.70\right)$ & $881$ & $18$ & $0.6 \pm 0.0$ & $1.1 \pm 0.3\%$ & $3.5 \pm 0.8\%$ \\
$\left[0.70,\ 1.00\right)$ & $2252$ & $46$ & $0.6 \pm 0.0$ & $1.3 \pm 0.2\%$ & $3.9 \pm 0.6\%$ \\
\cutinhead{star-forming, low mass ($10.6 < \logMsun < 10.9$) sample}
$\left[0.25,\ 0.45\right)$ & $225$ & $4$ & $1.4 \pm 0.2$ & $2.5 \pm 1.3\%$ & $7.5 \pm 4.0\%$ \\
$\left[0.45,\ 0.70\right)$ & $469$ & $10$ & $2.0 \pm 0.2$ & $4.3 \pm 1.5\%$ & $13.1 \pm 4.4\%$ \\
$\left[0.70,\ 1.00\right)$ & $660$ & $22$ & $3.4 \pm 0.3$ & $11.2 \pm 2.6\%$ & $33.9 \pm 7.9\%$ \\
\cutinhead{quiescent, low mass ($10.6 < \logMsun < 10.9$) sample}
$\left[0.25,\ 0.45\right)$ & $243$ & $2$ & $0.6 \pm 0.0$ & $0.5 \pm 0.3\%$ & $1.4 \pm 1.0\%$ \\
$\left[0.45,\ 0.70\right)$ & $393$ & $6$ & $0.6 \pm 0.0$ & $0.9 \pm 0.4\%$ & $2.6 \pm 1.1\%$ \\
$\left[0.70,\ 1.00\right)$ & $957$ & $18$ & $0.6 \pm 0.0$ & $1.2 \pm 0.3\%$ & $3.6 \pm 0.9\%$ \\
\cutinhead{star-forming, high mass ($\logMsun \geq 10.9$) sample}
$\left[0.25,\ 0.45\right)$ & $188$ & $4$ & $1.4 \pm 0.2$ & $3.0 \pm 1.6\%$ & $9.0 \pm 4.8\%$ \\
$\left[0.45,\ 0.70\right)$ & $294$ & $11$ & $2.0 \pm 0.2$ & $7.6 \pm 2.5\%$ & $23.0 \pm 7.5\%$ \\
$\left[0.70,\ 1.00\right)$ & $471$ & $14$ & $3.4 \pm 0.3$ & $10.0 \pm 2.8\%$ & $30.3 \pm 8.6\%$ \\
\cutinhead{quiescent, high mass ($\logMsun \geq 10.9$) sample}
$\left[0.25,\ 0.45\right)$ & $211$ & $5$ & $0.6 \pm 0.0$ & $1.3 \pm 0.6\%$ & $4.0 \pm 1.8\%$ \\
$\left[0.45,\ 0.70\right)$ & $488$ & $12$ & $0.6 \pm 0.0$ & $1.4 \pm 0.4\%$ & $4.2 \pm 1.2\%$ \\
$\left[0.70,\ 1.00\right)$ & $1295$ & $28$ & $0.6 \pm 0.0$ & $1.4 \pm 0.3\%$ & $4.1 \pm 0.8\%$ \\
\enddata
\tablecomments{The fraction of mergers is corrected by a factor of \Cmerge{} for line-of-sight superpositions, contamination from minor mergers/non-mergers, and  incompleteness. The fractional merger rate is calculated using $\langle\Tobs\rangle=0.33$ \citep{Lotz2011}.}
\end{deluxetable}
}

\ifthenelse{\boolean{astroph}}{
  \newcommand{\middfig}[1]{{#1}}
  \newcommand{\tailfig}[1]{{}}
}{
  \newcommand{\tailfig}[1]{{}}
  \newcommand{\middfig}[1]{{#1}} 
}

\newcommand{\myemail}{claire.lackner@ipmu.jp}

\shorttitle{Galaxy Mergers in COSMOS}
\shortauthors{Lackner et al.}

\begin{document}

\title{Late-stage galaxy mergers in COSMOS to $z\sim~1$}

\author{C.~N.~Lackner\altaffilmark{1}, J.~D.~Silverman\altaffilmark{1}, M.~Salvato\altaffilmark{2}, P.~Kampczyk\altaffilmark{3}, J.~S.~Kartaltepe\altaffilmark{4},  D.~Sanders\altaffilmark{5}, P.~Capak\altaffilmark{6}, F.~Civano\altaffilmark{7,8}, O.~Ilbert\altaffilmark{9}, K.~Jahnke\altaffilmark{10},  A.~M.~Koekemoer\altaffilmark{11},
N.~Lee\altaffilmark{5}, O.~Le~F{\`e}vre\altaffilmark{9}, C.~T.~Liu\altaffilmark{12}, N.~Scoville\altaffilmark{6}, K.~Sheth\altaffilmark{13}, and S.~Toft\altaffilmark{14}}
\altaffiltext{1}{Kavli IPMU (WPI), The University of Tokyo, Kashiwa, Chiba 277-8583, Japan}
\altaffiltext{2}{Max-Planck-Institut f\"ur extraterrestrische Physik, D-84571 Garching, Germany}
\altaffiltext{3}{Institute of Astronomy, Department of Physics, ETH Z\"urich, CH-8093 Z\"urich, Switzerland}
\altaffiltext{4}{National Optical Astronomy Observatory, Tucson, AZ 85719, USA}
\altaffiltext{5}{Institute for Astronomy, University of Hawaii, Honolulu, HI 96822}
\altaffiltext{6}{California Institute of Technology, Pasadena, CA 91125}
\altaffiltext{7}{Harvard-Smithsonian Center for Astrophysics, Cambridge, MA 02138}
\altaffiltext{8}{Department of Physics and Astronomy, Dartmouth College, Hanover, NH 03755}
\altaffiltext{9}{Aix Marseille Universit\'e, CNRS, LAM (Laboratoire d’Astrophysique de Marseille), 13388, Marseille, France
}
\altaffiltext{10}{Max-Planck-Institut f\"ur Astronomie, D-69117 Heidelberg, Germany}
\altaffiltext{11}{Space Telescope Science Institute, Baltimore, MD 21218}
\altaffiltext{12}{Astrophysical Observatory, CUNY, College of Staten Island, NY 10314 USA}
\altaffiltext{13}{National Radio Astronomy Observatory/NAASC, Charlottesville, VA 22903}
\altaffiltext{14}{Dark Cosmology Centre, Niels Bohr Institute, University of Copenhagen, Copenhagen, DK-2100 Denmark}
\email{\myemail}

\begin{abstract}
The role of major mergers in galaxy and black hole formation is not well constrained. To help address this, we develop an automated method to identify late-stage galaxy mergers before coalescence of the galactic cores. The resulting sample of mergers is distinct from those obtained using pair-finding and morphological indicators. Our method relies on median-filtering of high-resolution images in order to distinguish two concentrated galaxy nuclei at small separations. This method does not rely on low surface brightness features to identify mergers, and is therefore reliable to high redshift. 
Using mock images, we derive statistical contamination and incompleteness corrections for the fraction of late-stage mergers. The mock images show that our method returns an \rr{uncontaminated ($<10\%$) sample of mergers with projected separations between $2.2$ and $8$ \kpc{} out to $z\sim1$.} We apply our new method to a magnitude-limited ($m_{FW814} < 23$) sample of $44164$ galaxies from the COSMOS {\it HST}/ACS catalog. Using a mass-complete sample with $\log M_*/M_\odot > 10.6$ and $0.25 < z \leq 1.00$, we find $\sim 5\%$ of systems are late-stage mergers. Correcting for incompleteness and contamination, the fractional merger rate increases strongly with redshift as \rr{$\Rmerge \propto (1+z)^{3.8\pm0.9}$}, in agreement with
earlier studies and with dark matter halo merger rates. 
\rr{Separating the sample into star-forming and quiescent galaxies shows that the merger rate for star-forming galaxies increases strongly redshift, $(1+z)^{4.5\pm1.3}$, while the merger rate for quiescent galaxies is consistent with no evolution, $(1+z)^{1.1\pm1.2}$. The merger rate also becomes steeper with decreasing stellar mass.} Limiting our sample to galaxies with spectroscopic redshifts from zCOSMOS, we find that the star formation rates and X-ray selected AGN activity in \ed{likely} late-stage mergers are enhanced by factors of $\sim 2$ relative to a control sample. 
Combining our sample with more widely separated pairs, we find that $8\pm5\%$ of star formation and $20\pm8\%$ of AGN activity is triggered by close encounters ($<143\ \kpc$) or mergers, once more suggesting that major mergers are not the only channels for star formation and black hole growth.
\end{abstract}


%
\section{Introduction}
\label{intro}
In a hierarchical universe, galaxies grow by accretion of gas and mergers. Dark matter simulations suggest that the halo merger rate (in units per Gyr) increases with redshift as $(1+z)^{2-3}$ \citep{Lacey1993, Fakhouri2010}. However, identifying merging galaxies and transforming those observations into a galaxy merger rate is not easy, as is evidenced by the large differences between different methods \citep[see][and references therein]{Patton2008, dePropris2007, Bell2006, Jogee2009, Lotz2011}. Many of the differences in merger rates are due to differences in sample selection and merger identification \citep[see][]{Lotz2011}. Nonetheless, a precise determination of the merger rate of galaxies is essential to the study of galaxy growth. In particular, the galaxy merger rate is needed to compare the growth of galaxies to the growth of dark matter halos. Galaxy mergers may also play an important role in shaping galaxy morphology \citep{Toomre1972, Sanders1988, Barnes1992, Hopkins09}, instigating star formation \citep{Mihos1992, Sanders1996, Barton2000, Lambas2003, Ellison2008, Patton2013}, and inducing  super-massive black hole growth \citep[e.g.,][]{Hernquist1989, Moore96, Hopkins2008a, diMatteo2008}. 

There are two general classes of methods for finding galaxy mergers. The first class of methods selects close pairs of galaxies, before the galaxies have merged \rr{\citep[e.g.,][]{LopezSanjuan2013, LopezSanjuan2011, LopezSanjuan2012, Bundy2009, Lin2004, Lin2010, Patton2008, Kartaltepe2007,  deRavel2009, Bell2006, Masjedi08, Ellison2008, Williams2011, Robaina2010, Newman2012, Ellison2013, Tasca2014}}. These methods typically select galaxies with projected separations less than $100\,\kpchone$. The line-of-sight separation depends on the method used. Galaxy mergers selected using photometric redshifts include pairs that are widely separated along the line-of-sight and will never merge \citep[e.g.,][]{Kartaltepe2007, Bundy2009, LopezSanjuan2011, LopezSanjuan2012}. \rr{Although these methods cannot identify individual chance superpositions, superpositions can be easily accounted for statistically in merger-rate calculations \citep[e.g.][]{Bundy2009, LopezSanjuan2011, LopezSanjuan2012}.} In order to better distinguish chance superpositions from actual mergers, several studies utilize spectroscopic redshifts and identify kinematic pairs of galaxies \citep[e.g.,][]{Lin2004, Patton2008, deRavel2009, deRavel2011, Kampczyk2013, Ellison2013, LopezSanjuan2013}. For spectroscopic samples, understanding the completeness of the spectroscopy as a function of galaxy separation is essential \citep{Kampczyk2013, Patton2008, Lin2004}. Although kinematic pairs eliminate many chance superpositions from a sample of galaxy mergers, spectroscopic samples contain far fewer galaxies than photometric samples and often miss close pairs due to fiber collisions, leading to poorer statistics when measuring galaxy merger rates. 

The second class of methods for finding galaxy mergers looks for morphological signatures \ed{during or after} a merger (e.g., double nuclei, tidal tails, outer shells). Morphological searches either involve visual inspection \citep[e.g.,][]{Kartaltepe2014, Kartaltepe2010, Kampczyk2007, Jogee2009, Bridge2010, Cisternas2011}, or quantitative, non-parametric measures of galaxy morphology \citep[e.g.,][]{Lotz2008,Conselice2009,Scarlata2007,Shi2009,dePropris2007, LopezSanjuan2009}. Visual inspection involves searches for merger signatures, some obvious, such as double nuclei, and some subtle, such as sharp breaks in the radial light profile. Measurements, such as the the central concentration, asymmetry, and clumpiness (CAS) \citep{Abraham1994, Conselice2003}, the second moment of the light profile ($M_{20}$) \citep{Lotz2004}, and the Gini coefficient of the 2-dimensional flux distribution \citep{Abraham2003, Lotz2004}, seek to quantify the morphological signatures of bulges, disks, and galaxy mergers. By comparing visual classification with these quantitative measures, \citet{Lotz2004} demonstrate that major mergers occupy a distinct part of the Gini-$M_{20}$-concentration-asymmetry space. However, neither visual classification nor non-parametric morphology methods directly measure the merger mass ratio and do not distinguish between major and minor mergers. In fact, morphology-based methods are sensitive to very minor mergers and even close passages which cause morphological disturbances without leading to a merger \citep[see][]{Lotz2011}.

Both pair-finding methods and morphology-based methods are used to measure the galaxy merger fraction and its evolution. The methods are applied to many data sets with various selection functions. The resulting fractions of merging galaxies at $z\sim0.5$ vary from $1-20\%$ and the evolution of the merger rate is either very steep, $(1+z)^4$, or non-existent, $(1+z)^0$. Several works address these differences \citep[e.g.,][]{Patton2008, Jogee2009, Lotz2011}. Some of the differences may be due to the parent sample selection with redshift; samples selected at fixed number density show a different merger rate evolution than those selected at fixed stellar mass \citep{Lotz2011}. Additionally, the length of time each method is sensitive to a galaxy merger is highly uncertain and varies greatly as a function of merger-finding method and as a function of redshift and merger mass ratio. Both photometric pair-finding methods and morphological methods mis-identify galaxy mergers and often select chance superpositions of galaxies that are widely separated along the line of sight. \ed{Spectroscopic pair studies greatly limit the number of line-of-sight pairs, however.} Morphology-based methods often select galaxies in which close passages or very minor mergers caused dramatic morphological disturbances. While most studies correct for these mis-identifications, the correction factor is difficult to accurately calculate. The typical assumed fraction of mis-identified mergers ranges from $0$ to $60\%$ \citep[see][and references therein]{Lotz2011}.

In this work, we present a new quantitative method for identifying merging galaxies. Our method is in essence a high-pass filter which makes multiple peaks in galaxy surface brightness profiles easily distinguishable. The method is designed to select a sample of late-stage mergers in which two galaxy nuclei 
 are still intact and only separated by a few \kpc. In particular, we select galaxies whose nuclei are separated by $2.2-8$~$\kpc$, and expected to merge within a few hundred Myr \citep[see][]{Lotz2011}. These galaxies lie at the interface between early-stage galaxy mergers selected in close pair studies and post-merger galaxies selected based on disturbed morphologies. Indeed, we show below our sample of late-stage mergers has little overlap with kinematically selected, more widely separated pairs. At separations of a few \kpc, it is less likely that the galaxy pair is a chance superposition than at larger separations \ed{\citep[e.g.][]{Kartaltepe2007}}, reducing contamination from spurious pairs. 
By eliminating pairs that are likely to be spurious mergers, our estimates of the merger rate are robust and competitive with other merger-finding methods, including searches for spectroscopic pairs. While many galaxies in our sample could be selected by visual inspection, in practice there is little overlap between mergers identified by other quantitative morphology methods (e.g., Gini-$M_{20}$, asymmetry) and our sample of late-stage mergers. This is likely due to the fact galaxies with two very close, equally-bright central peaks will not have abnormally large asymmetry or second moment measurements. 

In addition to the number of merging galaxies, the properties of merging galaxies 
are also of much interest. Numerical simulations of merging galaxies demonstrate that major mergers can drive gas toward the center of galaxies leading to enhanced star formation, efficient bulge creation, and AGN activity as some of the gas is deposited on the central black hole \citep[e.g.,][]{Hernquist1989, Mihos1992, Barnes1992, Hopkins2008a, Hopkins2006}. Several observational studies have shown that merging galaxies typically have enhanced star formation \rr{\citep[e.g.,][]{Kartaltepe2012, Robaina2009, Xu2012, Kampczyk2013, Hung2013}}. Indeed, almost all intensely star-forming systems, such as luminous infrared galaxies (LIRGS), in the local universe have morphologies consistent with major mergers \citep{Sanders1996, Wu1998, Cui2001, Kartaltepe2010}. Similarly, several studies between $z\sim 0$ and $z\sim 1.2$ show an enhancement in the fraction of close galaxy pairs with AGN activity compared to isolated galaxies \citep[e.g.,][but see \citealp{Darg2010, Li2008, Ellison2008}]{Kennicutt1984, Keel1985, Alonso2007, Silverman2011, Koss2011, Ellison2013, Satyapal2014}. Nonetheless, the majority of low-luminosity AGN activity is not associated with merging galaxies \citep[e.g.,][]{Cisternas2011, Kocevski2012, Kauffmann2004,Silverman2011}. Using COSMOS {\it HST} images, \citet{Cisternas2011} find that X-ray selected AGN are no more likely to be ongoing or recent major mergers than a control sample of inactive galaxies. \citet{Silverman2011} examine \ed{close} kinematic pairs with separations \ed{less than $75 \kpc$} 
and find that $18\%$ of \ed{X-ray selected} AGN activity since $z\sim 1$ is \ed{triggered by interactions of} close pairs. Including late-stage mergers \ed{will increase the completeness of pairs at separations less than $10~\kpc$} and boost the fraction of AGN associated with merging. Late-stage mergers may help explain a significant fraction of the AGN activity not currently associated with galaxy pairs. Furthermore, understanding galaxy properties as a function of time to coalescence can help test numerical models of merging galaxies and black hole growth \citep[e.g.,][]{Hopkins2006, DiMatteo2007, Mihos1992}.

In this study, we apply our method, which selects galaxies likely undergoing late-stage mergers, to a flux-limited sample of $44164$ galaxies from the COSMOS survey. Because we are looking at high redshift and therefore require high spatial resolution to separate galaxy pairs, we apply our pair-finding method to {\it Hubble Space Telescope} ({\it HST}) images taken as a part of the COSMOS survey \citep{Scoville2007, Koekemoer2007}. The galaxies are selected from the ACS galaxy catalog \citep{Leauthaud2007}. Photometric redshifts, stellar masses, and absolute magnitudes are taken from the most recent near infrared-selected COSMOS catalog \citep{Ilbert2013, McCracken2012}. The data and merger-finding method are described in detail in \S\ref{sec:data} and \S\ref{sec:dblnuc}. The sample of $2047$ late-stage mergers is presented in Table \ref{tabp}. Thirty-two of these late-stage mergers are also detected in X-ray by either {\it Chandra} or {\it XMM}. 

We test the robustness of our method using simulated images of mergers in \S\ref{ssec:mocks}.  After demonstrating that our selection of late mergers is almost independent of redshift, we calculate the galaxy merger rate as a function of redshift between $0.25\leq z < 1.0$ (\S\ref{sec:mergerrates}). Finally, we analyze the star formation rates and \ed{X-ray selected} AGN fractions for the sub-sample of our data with spectroscopy from zCOSMOS \citep{Lilly2007, Lilly2009}. Unless otherwise noted, we use the cosmology $H_0=70\ \kmps\mathrm{Mpc}^{-1}$, $\Omega_m=0.25$, $\Omega_\Lambda=0.75$. \ed{When referring to other studies, we use units of $\kpchone$ where $H_0=100\ \kmps\mathrm{Mpc}^{-1}$. }
As in \citet{Ilbert2013}, magnitudes are given on the AB system \citep{Oke1974} and stellar masses in units of \Msun with a \citet{Chabrier2003} IMF.  

%
\section{Data}
\label{sec:data}
Finding late-stage galaxies mergers with two intact nuclei out to $z\sim~1$ requires the high resolution available in space-based images. We apply our pair finding method to {\it HST}/ACS F814W ($I-$band) images taken as part of the COSMOS project \citep{Koekemoer2007, Massey2010}. \ed{The pixel scale in these images in $0.03$\arcsec/pixel and the point spread function has a FWHM of $\sim0.09\ $pixels, which corresponds to $0.6$ \kpc{} at redshift $z=1$. Although our merger-finding method could be applied to ground-based data, it requires that stellar concentrations separated by a few \kpc{} are resolved. In {\it Sloan Digital Sky Survey} (SDSS) images, the median seeing in $1.3\arcsec$, which corresponds to $2.4$~\kpc{} at $z=0.1$. Therefore, our method could only be applied to SDSS data at $z<0.1$. For ground-based data with better seeing, the redshift limit could be increased to $z\sim0.3$. However, in order to study the evolution of the merger rate at significant redshift, {\it HST} data is necessary.}

For each galaxy in the parent sample, we create an $8\arcsec\times8\arcsec$ cutout from the ACS F814W image. These cutouts are used for detection of late-stage mergers. For this work we use two, overlapping samples of galaxies, both selected from the COSMOS ACS catalog \citep{Leauthaud2007}. The first sample uses photometric redshifts and contains $\sim 44,000$ galaxies. The second sample includes $\sim17,000$ galaxies and uses spectroscopic redshifts from the zCOSMOS project \citep{Lilly2007, Lilly2009}. Because of its greater size, we use the photometric sample to study the merger rate as a function of redshift. We use the spectroscopic sample to study the star formation and AGN activity in late-stage mergers. 

\subsection{Photometric Redshift Sample}
\label{ssec:photosample}
Our parent galaxy sample is selected from the COSMOS ACS catalog \citep{Leauthaud2007}. We select all objects classified as galaxies (\verb+MU_CLASS+ = 1) with total magnitudes brighter than $I=23$. In the case of a merger, the total magnitude of the post-merger galaxy will be brighter than $I=23$, while the individual components of the merging galaxy may be up to $5$ times fainter ($I \approx 24.7$), but still within the magnitude limit of the COSMOS ACS sample. The magnitude limit is necessary because the completeness of our merger-finding method decreases for lower signal-to-noise (see \S\ref{ssec:mocks}). 
We obtain redshifts and stellar masses for the galaxies in the ACS-selected catalog by matching to the recent $K$-band selected sample of COSMOS galaxies with photometric redshifts \citep{McCracken2012, Ilbert2013}. This catalog includes photometric redshifts and stellar masses for $90\%$ of the galaxies in our ACS-selected sample. The missing galaxies are due to slight differences in the area and the masking between the ACS and $K-$band catalogs. We exclude all galaxies masked in \citet{Ilbert2013}, as these galaxies do not have reliable photometric redshifts or stellar masses. \citet{Ilbert2013} report a photometric redshift precision of $\sigma_{\Delta z/(1+z)} = 0.008$ for $i^+ < 22.5$. The final sample contains \rr{$44164$} galaxies.

In \S \ref{ssec:agn}, we study the X-ray selected AGN fraction in late-stage mergers. Therefore, unlike \citet{Ilbert2013}, we include known X-ray sources in our parent galaxy sample.  For these sources, we use photometric redshifts from \citet{Salvato2011} for optical/NIR sources matched to {\it XMM} \citep{Brusa2010} and {\it Chandra} \citep{Civano2012} detections. \ed{Depending on the type of AGN, these photo-zs are computed using different galaxy-AGN hybrid templates, different luminosity priors, and accounting for source variability.} Note that the {\it Chandra} survey only covers $\sim 1/2$ of the COSMOS area, although at a greater depth.  Out of the parent sample of \rr{$44164$} galaxies, \rr{$502$} galaxies have an {\it XMM} counterpart, and \rr{$573$} have a {\it Chandra} counterpart, with \rr{$282$} sources detected by both instruments.  For X-ray sources, the photo-zs have a precision of $\sigma_{\Delta z/(1+z)} = 0.015$ \citep{Salvato2011}. For the sources identified in {\it XMM}, we use stellar masses computed by \citet{Bongiorno2012}. These masses and photo-zs are most reliable for galaxies that are not AGN/quasar dominated (type I). For the analysis of the AGN fraction, we will restrict our galaxy sample to systems that are not AGN-dominated, based on the photo-z templates used. This will eliminate galaxies with the least certain photo-zs and stellar masses.

When matching the ACS catalog with the ground-based $K$-band selected catalog, $1\%$ of sources ($640$ sources) that are resolved into $2$ galaxies in the {\it HST} data are not resolved in the ground-based data. Clearly, these galaxies are possible late-stage mergers with small separations, and, therefore, cannot simply be removed from the sample. For these cases, we ensure that the galaxy is only included in the sample once, and we use the sum of the $I$-band magnitudes for the total magnitude. \rr{Because our sample is selected from the ACS catalog, we may be missing resolved late-stage mergers in which both of the components are below the $I-$band magnitude limit. However, resolved galaxy pairs make up less than $10\%$ of our final late-stage merger sample, as most of the resolved pairs are more widely separated than the late-stage mergers selected below.} From visual inspection, $\sim 70\%$ of these resolved galaxy pairs show clear signs of interaction, while the remaining pairs may be chance superpositions. This fraction of chance superpositions agrees well with that obtained by other methods (see \S \ref{ssec:contaim}). However, by randomly adding galaxies to COSMOS images, \citet{Kampczyk2007} find the fraction of chance superpositions in a sample of visually-selected mergers is $40\%$ at $z\sim0.7$, suggesting that visual inspection cannot reliably distinguish real mergers from chance superpositions. Because we use visual inspection to ascertain the number of chance superpositions, we may underestimate the fraction of chance superpositions by almost a factor of two. 

For chance superpositions in which two galaxies at different redshifts are unresolved by ground-based observations, the photometric redshifts are especially suspect \citep{Bordoloi2010}. Indeed, in comparing the photometric redshifts to the available spectroscopic redshifts (see \S\ref{ssec:specsample}), we find the rate of catastrophic outliers for possible late-stage mergers is $1.6\%$ compared to $1.0\%$ for the entire sample of galaxies between $0.25<z_{\mathrm{spec.}}<1.0$. However, the catastrophic outlier rate for the late stage merger candidates is still small, and the precision of the photometric redshifts remains unchanged ($\sigma_{\Delta z/(1+\mathrm{specz})}=0.004$). In this work, we will assume the photometric redshifts of the late-stage merger candidates are as accurate as those for the entire sample. The comparison to spectroscopic redshifts does not take into account that the spectroscopic redshifts of late-stage mergers may also be suspect, because the galaxies are typically separated by less than an arcsecond and their spectra are blended. We have visually inspected the spectra of late-stage mergers and do not find any in which two merging galaxies are easily discernible in ground-based spectroscopic data.

\subsection{Spectroscopic redshift sample}
\label{ssec:specsample}
In \S\ref{sec:props}, we compare the AGN and star formation rates in our sample of late-stage mergers to kinematically-selected pairs \citep{Kampczyk2013, Silverman2011} from the zCOSMOS survey \citep{Lilly2007, Lilly2009}. For the study of SFR (\S\ref{ssec:sfr}), we use the bright zCOSMOS 20k bright sample ($I < 22.5$), which contains $16467$ galaxies with reliable redshifts. Following \citet{Lilly2009}, we only use galaxies with redshifts in the confidence classes: 1.5, 2.4, 2.5, 3.x, 4.x, and 9.5, as well as secondary targets with the same redshift confidences. 
In principle, the spectroscopic sample should be an exact subset of the larger ACS-selected photo-z sample above. However, differences in masking and the observed region exclude $\sim 1200$ galaxies in the zCOSMOS survey from the photometric redshift sample described above. \ed{Unlike kinematic pair studies in which both members of the merger have measured spectroscopic redshifts, in our sample of late-stage mergers, we only have one spectroscopic measurement for the entire merging system. Most of the late-stage merger candidates described below are separated by less than an arcsecond, so their spectra are blended. As noted above, we have visually inspected a subset of the spectra for these galaxies, and do not find any cases in which two sets of lines (two redshifts) are easily discernible. }

In \S\ref{ssec:agn}, we study the fraction of X-ray selected AGN in late stage mergers compared to the fraction of AGN in kinematic pairs. In order to make the comparison straightforward, we use the same parent sample as defined by \citet{Silverman2011} and \citet{Kampczyk2013}. As in \citet{Silverman2011}, we use {\it Chandra} observations to identify AGN. Because the {\it Chandra} survey \citep{Elvis2009} only covers $\sim1/2$ of the COSMOS field,  the parent sample of galaxies is smaller. Therefore, we only examine $10681$ galaxies from the bright zCOSMOS survey which lie within the {\it Chandra} footprint.

Table \ref{tabsamps} lists the various parent samples as well as the their properties. We have also included the cuts made to these samples for the analysis in \S\ref{sec:mergerrates} and \S\ref{sec:props}.
\tablesamps

\section{Merging galaxy selection}
\label{sec:dblnuc}
 To separately detect each component in a merging galaxy, we run the images through a high-pass filter, which makes multiple peaks in the flux distribution easily distinguishable. Our procedure, illustrated in Figure \ref{fig:demo}, is as follows:
\begin{itemize}
\item We first convolve each postage stamp image with a median ring filter \citep{Secker1995}. This smooths the image by replacing each pixel with the median value in a ring surrounding that pixel, thus erasing structures on scales larger than the ring. We set the inner ring radius to $9$ pixels, which is approximately $3$ times the PSF width. This sets the size of the smallest separation we can detect. At $z\sim 1.0(0.2)$, $9$ pixels corresponds to $2.2(1.1)$ \kpc. For comparison, in SDSS images at $z\sim 0.1$, a $9$-pixel radius median ring filter could only detect peaks separated by at least $6$ \kpc. \ed{In order to apply this method to ground-based data, the size of the median ring filter (in pixels) would have to be adjusted for the seeing.}
\item We then subtract the smoothed image from the original image. Together, the first two steps create a high-pass filter.
\item In the difference image, we select all pixels $5$ standard deviations above the noise. Contiguous regions are considered a single peak. For a peak detection to be significant, we require a region to contain at least $8$ pixels. These values ensure that any detected peak is at least as large as the PSF. We demonstrate in \S\ref{ssec:mocks} that using these detection thresholds yields a sample of peaks which is relatively complete and uncontaminated, \rr{particularly for mergers between early-type, bulge-dominated galaxies}.
\end{itemize}
\middfig{\figdemo}

Many of the detections returned by this algorithm are not actually merging galaxies, but rather widely separated galaxies (our postage stamp images span $>50\kpc$ at $z\sim1$), galaxies with clumpy star formation, spiral arms or bars. \ed{Another source of contamination is edge on disk galaxies in which a large bulge is bisected by the dusty disk.} In order to eliminate most of these spurious mergers, we place further restrictions on the detected peaks. 

First, we require that the peaks are separated by no more than $8$ \kpc. We use this upper limit to restrict the sample to galaxies that are likely to be mergers, not just close pairs or chance superpositions. Additionally, this eliminates the problem of double-counting late-stage, since widely separated galaxies will be separate detections in the parent galaxy sample. 

In order to study the fraction of merging galaxies as a function of redshift, we also implement a lower bound on the peak separation of 2.2 \kpc. This lower limit is set by the size of the median ring filter. In this way, we are sensitive to the same mergers at both low and high redshift. Implementing this lower bound on the peak separation eliminates \rr{$25\%$} of the late-stage mergers, mainly at low redshift.

Second, we remove detected peaks that are faint compared to the brightest peak in each galaxy \emph{and} to the galaxy as a whole.  We measure the flux in each component simply by summing the flux in the pixels associated with the peak. Because this only includes the flux in the central region of each merging component, this is an underestimate of the flux in each component of a merger. \rr{Based on merger selection in mock images (see \S\ref{ssec:mocks} and Appendix \ref{app:sims}), we require that every detected peak contains at least $3\%$ of the the total flux in the galaxy. We demonstrate below that this successfully eliminates $80-90\%$ of the contamination from non-merging galaxies and star-forming clumps, without greatly affecting the completeness of our sample.  In order to only study major mergers, we require that every detected peak is at least $1/4$ as bright as the brightest peak detected for each galaxy. However, while this cut helps eliminate contamination from minor mergers (see Figure \ref{fig:fcontaim}), the measured flux ratio is inaccurate as are cuts based on it. Together, the cuts in flux ratio eliminate $75\%$ of the galaxies that would otherwise be considered late-stage mergers. However, neither of these cuts significantly affects our completeness, which is determined by the efficacy of the median ring filter. After these cuts, our overall completeness is $\sim 20\%$, but this increases to $\sim80\%$ for mergers between bulge-dominated galaxies.}


Finally, there are some images that have more than $2$ detected peaks. We expect triple merging systems to be extremely rare, and visual inspection shows that most images with $3$ or more peaks, after removing faint peaks, are edge-on disk galaxies, in which the bulge and the ends of the spiral arms or bar are detected. These galaxies can be eliminated from the late-stage merger sample by requiring that multiple peaks do not lie along a single line, as is the case for edge-on galaxies. We implement this cut by requiring that the absolute value of the Pearson correlation coefficient for the peak centroids is less than $0.5$. \rr{After all the other cuts have been applied, this cut eliminates $7\%$ of the detected late-stage mergers ($145$ of $2047$). }

\middfig{\figexample}
\middfig{\figreject}
With these restrictions, we find \rr{$2047$} (\rr{$1547$} with separations greater than $2.2\ \kpc$) late-stage mergers with two prominent flux peaks in the photo-z sample of \rr{44164} galaxies. These are listed in Table \ref{tabp} along with some basic properties of the galaxies and the detected peaks. In the spectroscopic sample of \rr{$16467$} galaxies, we find \rr{$819$} merging galaxies, \rr{$71$} of which are not included in the photo-z sample due to differences in masking. The late-stage mergers in the zCOSMOS sample are also listed in Table \ref{tabp}. For each late-stage merger, we include the projected separation between the two flux peaks as well as the flux ratio of the peaks. In the photo-z sample, \rr{$32$} mergers are X-ray AGN detected in either {\it Chandra} or {\it XMM}. In the spectroscopic redshift sample, \rr{$10$} late-stage mergers are matched with a {\it Chandra} source out of \rr{$534$} mergers that lie within the {\it Chandra} footprint. 

Although we are confident that the majority of late-stage mergers listed in Table \ref{tabp} are physical late-stage mergers, not all the systems identified by our method will be real mergers. In addition to isolated galaxies with clumpy central structure, our sample contains line-of-sight superpositions. Below, we show these chance superpositions represent $30\%$ of the late-stage merger sample. Without two spectroscopic redshifts for each member of the merger or detailed kinematic maps, it impossible to distinguish real late-stage mergers from widely-separated chance superpositions or from isolated galaxies with complex structures. Indeed, by tuning the selection criteria to accept smaller peaks, our median-ring filter method could be used to find galaxies with several bright clumps instead of late-stage binary mergers.
 
It is important to note that, other than the pair separation and flux ratio, all other measured properties (e.g., color, redshift, stellar mass, X-ray flux) are properties of the merger, not the individual component galaxies. If we divide galaxies by stellar mass, merging galaxies are counted with galaxies more massive than either member of the merger. In this way, late-stage mergers are treated more like post-merger galaxies than like pairs of galaxies in the early stages of merging. This distinction is important to keep in mind when comparing to samples of paired galaxies, in which properties for the individual galaxies are reported. 

Figure \ref{fig:example} shows images of $6$ late-stage mergers in the photo-z sample. Although the galaxy in the lower middle panel may be a spiral galaxy without any merging activity, the remaining galaxies are clearly mergers at various separations. The typical peak separation is less than $1\arcsec$, demonstrating why our algorithm requires the high resolution of space-based data. Figure \ref{fig:reject} shows three examples of galaxies which do not satisfy the cuts on peak flux ratio, peak separation, or peak Pearson correlation coefficient. These galaxies are often spiral or barred galaxies. 
As noted above the median ring filter detects edge-on disk galaxies, in which the bulge is bisected by dust in the disk, as binaries mergers. However, these galaxies represent a small contamination. 
At fixed star formation rate, the fraction of late-stage mergers candidates is independent of galaxy ellipticity (proxy for disk inclination), suggesting our detection algorithm is not biased by galaxy inclination and dust obscuration.

\middfig{\tablepairs}

\subsection{Simulated Merger Images}
\label{ssec:mocks}
\rr{
We test our merger detection algorithm on a set of simple mock images of merging galaxies. We create postage stamps of pairs of galaxies using the {\it HST}/ACS images from the photo-z sample. Each mock image contains $2$ random galaxies at the same photometric redshift. We choose galaxies at the same redshift in order to eliminate line-of-sight chance superpositions, which we address statistically  in \S\ref{ssec:contaim}. By using real galaxy images drawn at random, we can create a sample of mergers with realistic morphologies, magnitudes, and flux ratios. Note, however, that this method does not include any structural changes caused by mergers. We simply superimpose images of isolated galaxies.

For each pair of galaxies, we make $8$ mock images in which the galaxies are separated by  $0.5$, $1.5$, $2.0$, $2.5$, $3.0$, $5.0$, $7.0$, and $10.0$~\kpc{}. This allows us to test the completeness of our sample as a function of projected separation. Details of the mock images are discussed in Appendix \ref{app:sims}. Because the flux limits and separation limits used in this study are derived from these simulations, applications of the median-ring filter method to other data sets require new simulations matched to the observations and adjustments to the flux and separation limits given in \S\ref{sec:dblnuc}.

After running these mock merger images through the median ring filter, we examine the completeness of our merger sample as a function of pair separation, redshift, and flux ratio. Unsurprisingly, we find that the efficiency of detecting late-stage mergers drops precipitously if the pair separation is smaller than $\sim 10$ pixels, or the size of the median ring filter (see Figure \ref{fig:scomplete}). This motivates using a lower bound of 2.2 \kpc{} ($9$ pixels at $z\sim 1$) for the pair separation. The median ring filter selects $\sim 40\%$ of the mock mergers with separation larger than 2.2 \kpc{} to $z=1$. This completeness depends strongly on galaxy morphology. Because the median ring filter smooths away diffuse structures, our merger-finding method is biased toward mergers between concentrated, early-type galaxies\footnote{We use the ZEST parameter \citep{Scarlata2007} to ascertain the morphologies of the galaxies in a mock merger. See \S\ref{ssec:othertechs} for more details.}. For these mergers, our method is $80\%$ complete, while for mixed mergers (late$+$early) and mergers between late-type galaxies, the median ring filter only selects $40\%$ of mock mergers. After removing contamination, the completeness of our selection of early-type mergers is independent of redshift. For late-type galaxies, the completeness drops slightly at higher redshifts (see Figure \ref{fig:zcomplete}, right panel). For all morphologies together, the completeness drops between $z\approx0.2$ and $z\approx0.5$, as the fraction of early-type galaxies also decreases toward higher redshift.

In addition to using the mock mergers to study completeness, we can use them to study the contamination from non-merging, clumpy galaxies and minor mergers (see Figure \ref{fig:fcontaim}). We find that, unlike the completeness, the contamination is essentially independent of redshift. This may reflect the fact that our merger-finding method is less sensitive to all structures at lower signal-to-noise, and, therefore, higher redshift. Using artificially redshifted mergers, \citet{Kampczyk2007} find that mergers identified in low-z data will not appear as mergers at higher redshift due to lower resolution and signal-to-noise, which may explain the incompleteness of our merger selection at high redshift. However, unlike \citet{Kampczyk2007}, our merger identification does not take into account morphological k-corrections. Because galaxies are less smooth at bluer wavelengths and have a larger fraction of their flux concentrated in smaller regions, our peak-finding method may detect more non-merging, star forming galaxies at bluer rest frame wavelengths, increasing the contamination. Even if morphological k-corrections are taken into account, galaxies at high redshift are expected to be clumpier \citep[e.g.][]{Bournaud2007, Genzel2011} and have higher star formation rates, thus again increasing the contamination from non-merging, clumpy star-forming galaxies. The lack of redshift dependence in the contamination suggests our method is not particularly sensitive to clumpy star-formation at high redshift, possibly because these clumps are too small and faint. A better understanding of the effects of morphological k-corrections will require further study using multi-wavelength data, in particular near-infrared data at $z\sim1$.

We use the mock mergers to determine the cuts on peak flux to galaxy flux ratio ($>3\%$) and peak to peak flux ratio ($>25\%$). By implementing these cuts, we are able to reduce the contamination from non-merging galaxies to $10\%$, and the contamination from minor mergers (flux ratios smaller than $1$:$4$) to $20\%$. The median ring filter is naturally less sensitive to minor mergers than major mergers since the faint member of the merger is likely to be below our detection threshold. These cuts do affect the completeness, decreasing it by a factor of $2$ to $20\%$ (see Figure \ref{fig:zcomplete}). However, the completeness for early-type galaxy mergers is larger unaffected.  Removing the cut on peak flux to galaxy flux ratio increases the contamination from non-merging sources to $\sim 40\%$, which significantly affects our results on internal late-stage merger properties (SF, AGN), which cannot be simply corrected. In calculating the merger rate (\S\ref{sec:mergerrates}), we correct the measured late-stage merger fractions for contamination and incompleteness. For all mergers, we take the contamination to be $30\%$, independent of redshift. The incompleteness correction is a function of redshift and merger type and is derived from Figure \ref{fig:zcomplete}.
}

\subsection{Caveats}
\label{ssec:caveats}
Although our simulations demonstrate the effectiveness of our merger-finding, there are several failure modes of the algorithm. First, the method does not distinguish between merging galaxies and chance superpositions. Since we are looking at extremely small separations, we expect the number of chance superpositions to be small. Although we correct the merger rates in \S\ref{sec:mergerrates} for chance superpositions, we cannot correct the properties of mergers for contamination from chance superpositions. 

\rr{Second, tests with mock merger images show that the median ring-filter is most sensitive to highly concentrated galaxies. This suggests the merger rate measured for early-type (quiescent) galaxies is very robust, but the merger rate for late-type (star-forming) galaxies is underestimated by as much as a factor of $3$, accounting for contamination from non-mergers and line-of-sight superpositions. Furthermore, the bias toward highly concentrated galaxies may introduce biases in the sample as a function of redshift. Morphological k-corrections will lead to more disk-dominated, less centrally concentrated, galaxies at high redshift \citep{Kuchinski2000, Papovich2003}. Therefore, a bias toward detecting concentrated galaxies is likely to under-report the number of late-stage mergers at high redshift. }

Finally, because we impose a flux ratio cut, our method is not sensitive to a merger with only one optically bright AGN. Therefore, in \S\ref{ssec:agn}, we only compute the AGN fraction for obscured (Type II) AGN. Additionally, the stellar masses and photometric redshifts for Type I AGN are less certain than for obscured, optically-faint AGN. Thus, removing Type I AGN from the sample improves the accuracy of our results. 

\subsection{Comparison with other selection techniques}
\label{ssec:othertechs}
There are many established methods for selecting merging galaxies. The simplest methods select galaxies based on angular separation \citep[e.g.,][]{Carlberg1994, Zepf1989}. These methods typically look at separations of $5-100\,\kpchone$, and need to be corrected for chance superpositions and galaxies that are physically close, but will not merge within a Hubble time. The number of superpositions can be limited by requiring the galaxies have similar photometric redshifts \citep[e.g.,][]{Kartaltepe2007, Bundy2009}. Similarly, spectroscopic redshifts are also useful in eliminating chance superpositions \citep[e.g.,][]{Patton2002, Patton2008, Kampczyk2013, Lin2004, Lin2008, LeFevre00, deRavel2009, deRavel2011, Ellison2008}.  However, spectroscopic samples are limited in size and depth. Furthermore, the late-stage mergers reported here typically have sub-arcsecond separations. Systems with such small separations will not be resolved in ground-based spectroscopic studies, even when including pairs that are observed in the same slit \citep[e.g.][]{Kampczyk2013}.
Comparing our results to the spectroscopically-selected pair sample in \citet{Kampczyk2013}, we find only $20\%$ of late-stage mergers are also considered kinematic pairs \ed{and most of these mergers have separations larger than $2\arcsec$.} 

Pair samples look for galaxies in the early stages of merging. Morphological studies, on the other hand, look for evidence of \ed{mergers at all stages, including late-stage mergers and post-merger galaxies.} Merger studies based on morphology rely on visual classification \citep[e.g.,][]{Kampczyk2007, Bridge2010, Darg2010, Kartaltepe2010}, or quantitative morphology indicators. Indicators used to distinguish mergers include the Gini coefficient\footnote{The Gini coefficient measures relative distribution of flux in pixels associated with a galaxy. It is given by $G=1/\left(2\bar{f}n(n+1)\right)\Sigma_{i=1}^n\Sigma_{j=1}^n\left|f_i-f_j\right|$, where $f_i$ is the flux in a pixel, $n$ is the number of pixels, and $\bar{f}$ is the mean flux per pixel \citep{Abraham2003}} and the second moment of the brightest pixels, $M_{20}$\footnote{$M_{20}$ is the second moment of the flux around a galaxy's center ($\Sigma_{i=1}^nf_i(x_i^2+y_i^2)$), only counting the brightest pixels which total $20\%$ of the galaxy's flux. This is then normalized by the galaxy's total second moment, summing over all pixels \citep{Lotz2004}.} \citep[e.g.,][]{Abraham2003, Lotz2004, Lotz2008}; the galaxy asymmetry and concentration \citep[e.g.,][]{Conselice2009, Shi2009, LopezSanjuan2009}; and combinations of the above as well as parametric fits to galaxy luminosity profiles \citep[e.g.,][]{Scarlata2007, Cassata2005}.  
In particular, \citet{Scarlata2007} use principle component analysis to reduce the space spanned by Gini, $M_{20}$, concentration, asymmetry, clumpiness, and galaxy S\'ersic index to three dimensions. Regions in this space are then assigned a ZEST (Zurich Estimator of Structural Types) type. This estimator has been applied to the COSMOS ACS images, making comparisons to our late-stage merger sample possible. 
\middfig{\figginiM}
\middfig{\figconcenasym}

Figures \ref{fig:giniM20} and \ref{fig:concenasym} show the ZEST classifications of our sample. All morphological values are taken from \citet{Scarlata2007}.  As described in \citet{Scarlata2007}, the ZEST types are computed based on `clean' ACS images, in which close companions, if found, are masked. Therefore, we expect that for some mergers, one member of the pair will be masked, decreasing the measured asymmetry and $M_{20}$. \ed{However, inspection of the COSMOS `clean' images shows that the majority of late stage mergers are considered a single system \citep[see][]{Cisternas2011}.} We limit the sample to galaxies with stellar masses above $2.5\times 10^{10}\, \Msun$ and $I-$band magnitudes brighter than $23.5$. The mass-restricted parent sample is shown by colored contours, while the late-stage mergers are denoted by black points. In these figures, it is clear that most  mergers are either ZEST type 2 (bulge+disk galaxies) or ZEST type 3 (irregular). Out of $212$ late-stage mergers, $28$, $112$, and $72$ are of types 1, 2, and 3, respectively. If the late-stage merger sample had the same distribution as the parent sample, the expected number of each type would be $50$, $152$, and $10$. These differences in ZEST type distribution are shown in the histograms in Figures \ref{fig:giniM20} and \ref{fig:concenasym}. 

These histograms show the distributions of $M_{20}$ and concentration for late-stage mergers compared to normal galaxies for each ZEST type. For late-stage mergers classified as spirals, the distribution of $M_{20}$ is shifted toward larger values (blue lines), closer to the distribution for irregular galaxies (green line). This demonstrates that, while the ZEST categorization does not clearly separate late-stage mergers from spiral galaxies, the morphologies of late-stage mergers are measurably different from those of regular spiral galaxies. Nonetheless, most mergers are classified as spirals, in agreement with \citet{Kampczyk2007}. Furthermore, most irregular galaxies are not classified as mergers by our method. This is to be expected, since our merger selection only finds major mergers which still have two nuclei, ignoring other merger signatures.

In Figure \ref{fig:giniM20}, the magenta dashed line shows the criterion for merging galaxies from \citet{Lotz2004}. This criterion is designed for observations in the rest frame $B-$band and is therefore appropriate for the portion of our sample  above $z\approx0.7$. Objects above this line are considered mergers, However, only a small fraction of late-stage mergers are also mergers based on the G-$M_{20}$ criterion \citep[see also][]{Kartaltepe2010, Jogee2009, Lotz2011}. Similarly, in Figure \ref{fig:concenasym}, major mergers are expected to be highly asymmetric (asymmetry $\gtrsim 0.3$), but only a small fraction of our sample of late-stage mergers have such high asymmetries. 

\rr{In Appendix \ref{app:casgini}, we further explore the differences between our sample of late-stage mergers and mergers selected based on their Gini, $M_{20}$, or asymmetry values. Some differences may be due to the fact that the  Gini, $M_{20}$ and asymmetry values reported by \citet{Cassata2005} are based on deblended images, which will split two concentrated nearby galaxies into separated sources. This will eliminate late-stage mergers from the $G-M_{20}$ and asymmetry-selected samples. Furthermore, the high central concentrations of late-stage mergers selected here biases the asymmetry upward, making it less likely they are selected as mergers based on asymmetry. In looking at mergers selected by Gini-$M_{20}$ and asymmetry but \emph{not} by our method, we find that the asymmetry and Gini-$M_{20}$ methods are more sensitive to minor mergers and small perturbations in the galaxy flux distribution than our methods.  Together, these reasons help explain the poor overlap between our sample of late-stage mergers and those derived using other morphology methods.}

\section{Merger Rates}
\label{sec:mergerrates}
In section \S\ref{ssec:mocks}, we demonstrate \rr{the completeness and contamination of our selection of late-stage mergers. By correcting for these effects, we can compute the major merger rate as a function of redshift to $z=1$.} To calculate merger rates, we use the photo-z parent sample, with a few additional restrictions. We restrict our parent sample to the approximately volume-limited sample between $0.25\leq z < 1.0$ and $\log M_*/\Msun > 10.6$. The stellar mass limit is derived by comparing the completeness of our $I-$band selected catalog to that of the deeper $K-$band selected catalog \citep{Ilbert2013}. \rr{Up to $z=1$, $93\%$ of the galaxies from the deeper $K-$band selected catalog are included in our sample.  Figure \ref{fig:masscomp} shows the measured stellar masses as a function of apparent magnitude in $3$ redshift bins. While the sample is mass-complete for the lower redshift bins, the completeness  drops to $82\%$ for $z > 0.9$.} \rr{For the merger rate analysis, we also remove all sources with X-ray detections because the colors and photometric redshifts for these sources are less certain. This eliminates $3\%$ of the sample. } The final sample for the merger rate analysis contains \rr{$5894$} galaxies, of which \rr{$136$} are late-stage mergers. 
\middfig{\figmasscomp}

\subsection{Line-of-sight pairs correction}
\label{ssec:contaim}
Before computing the merger rate, we need to correct the observed number of late-stage mergers for chance superpositions. For each galaxy in the sample, we compute the expected number of projected neighbors by summing the average sky density of possible neighbors over the area of the search annulus between $2.2\ \kpc$ and $8\ \kpc$. This method is outlined in \citet{Bundy2009}. If a galaxy in the parent sample is identified as a merger, it is made of two galaxies with a flux ratio between 1:1 and 1:4. Therefore, for each galaxy the possible companion galaxies are those with fluxes between $1/2$ and $1/5$ the galaxy's flux. Using the deeper, full ACS source catalog \citep{Leauthaud2007}, we compute the average sky density of possible projected neighbor galaxies. The expected number of chance superpositions for each galaxy is simply the sky density of possible neighbor galaxies multiplied by the area where we search for mergers, namely an annulus with and inner(outer) radius of $2.2$~\kpc{}($8$~\kpc) around each galaxy in the parent sample. Summing the expected number of projected neighbors over the whole parent sample yields an expected number of chance super-positions of $41.4$. In other words, $30\%$ of the $136$ late-stage mergers are likely to be chance superpositions. \ed{Because the angular diameter distance, and, therefore, the size of the annulus searched for close pairs, changes slowly with redshift beyond $z\sim0.5$, the fraction of chance superpositions does not change significantly with increasing redshift.} Therefore, we correct the fraction of late-stage mergers by a factor of $\Clos=0.7$. This correction is only statistical and does not allow us to determine which late-stage mergers are chance superpositions, only the average probability that any late-stage merger is a spurious pair.

The value of $\Clos$ given above agrees well with the fraction of chance superpositions found by visually inspecting a fraction of the merger images in \S\ref{ssec:photosample}. Our correction factor also agrees well with other values for \Clos{} based on numerical simulations where $\Clos\approx0.4-1.0$ \citep[see][]{Kitzbichler2008, Patton2008, Lotz2011, LeFevre00} and visual inspection \citep{Kampczyk2007}. As expected, the fraction of chance superpositions in our sample is smaller than that found by \citet{Bundy2009} using the same method but a larger search annulus. By including photometric redshift information for both members of a merger, \citet{Kartaltepe2007} find a smaller correction factor ($\Clos \sim 15\%$). However, our method cannot distinguish the redshifts of the superimposed galaxies. Despite good agreement with previous studies, the value of \Clos{} remains highly uncertain and contributes significantly to the uncertainty in the merger rates calculated below.

\subsection{Contamination and completeness}
\label{ssec:contandcomp}
\rr{Using the mock merger images from \S\ref{ssec:mocks}, we correct our sample of late-stage mergers for incompleteness and contamination. The simulations show that the sample has a contamination rate of $33\pm1\%$, $\sim20\%$ from minor mergers and $\sim10\%$ from non-merging, clumpy galaxies. The contamination does not depend strongly on redshift, although it does depend on the flux ratio of the merger. For the merger rate calculation, we correct the number of late-stage mergers for contamination by multiplying by $0.67\pm0.01$. 

The simulations also demonstrate the incompleteness of our merger finding method. From Figure \ref{fig:zcomplete}, the completeness of our sample ranges from $\sim20\%$ to $\sim40\%$ as a function of redshift. Based on the mock mergers, we compute the completeness in the three redshift bins used below. The correction factor for incompleteness is simply the inverse of the completeness fraction. In addition to redshift, the completeness of our late-stage merger selection depends strongly on galaxy morphology. In \S\ref{ssec:mcolor}, we compute the merger rates for star-forming and quiescent galaxies separately. In this case, we use completeness corrections derived from the late-type and early-type galaxy mergers, respectively. While obtaining correct morphologies for artificially superimposed galaxies is trivial, obtaining correct colors requires understanding how the components of the mock merger extinct each other. Instead, we assume that the color (star formation rate) of a merger and the morphology of the merging galaxies are exactly correlated, i.e., there are no star-forming early-type mergers. This is certainly not true, but is likely a small error in light of the overall uncertainties in the corrections for incompleteness and line-of-sight chance superpositions.

Taken with the line-of-sight pair correction, we correct the number of late-stage mergers by three factors, \Clos{}, the contamination ($0.67\pm0.01$), and the incompleteness $1/\mathrm{completeness\ fraction}$. Together, we denote these corrections as \Cmerge{}. The values of \Cmerge{} are given in Table \ref{tabMR} along with the \emph{corrected} merger fraction \fecp{}. The errors on \Cmerge{} are the bootstrap-derived errors on the completeness and contamination fractions (see Appendix \ref{app:sims}) and do not include the uncertainty in \Clos{}. The redshift dependence of \Cmerge{} is due entirely to the incompleteness correction. In \S\ref{ssec:nocorrect}, we examine the measured merger rate without corrections for incompleteness and contamination in order to determine how sensitive our results are to these correction factors.}


\subsection{Evolution of the merger rate}
\label{ssec:evolmergerrate}
In order to calculate the merger rate, we simply count the number of pairs in three redshift bins, chosen such that each redshift bin spans the same amount of time. Our results are unchanged if the bins are chosen such that they contain the same number of galaxies. The raw merger fractions are corrected for chance superpositions, contamination, and incompleteness using the correction factor, \Cmerge{}, given in Table \ref{tabMR}. 
The corrected merger fraction for our total sample is \rr{$\Cmerge\times136/5894 = 4.8\pm 0.5\%$, where $\Cmerge=2.1\pm0.1$, and the completeness fraction is $0.22\pm0.01$.} This is comparable to the typical pair fraction found in studies of more widely separated pairs \citep{Lin2004, Kartaltepe2007, deRavel2009, Bundy2009, Robaina2010}. 
and the fraction of morphologically disrupted systems 
\citep{dePropris2007, Conselice2009, Lotz2008}. 
\rr{As noted in \S\ref{ssec:othertechs}, our sample of late-stage mergers has little overlap with samples of major mergers selected by visual inspection or other non-parametric methods (asymmetry, etc.). This makes the remarkably good agreement in the overall fraction of mergers difficult to interpret, and possibly due to chance. }
\middfig{\tablemr}

In order to compute the galaxy merger rate, we must take into account the timescale over which a late-stage merger could be observed. We calculate the fractional merger rate as it is defined in \citep{Lotz2011}:
\begin{equation}
\Rmerge = \fecp\langle\frac{1}{\Tobs}\rangle\, ,
\end{equation}
where \Tobs{} is the duration of time a merger will be observable. Because \Tobs{} is highly dependent on the merger/pair selection, using the correct value for \Tobs{} is essential when comparing merger rates based on different techniques \citep[see][]{Lotz2011}. For the late-stage mergers studied here, \Tobs{} is sensitive to many parameters such as the galaxy masses, gas fractions, orbital parameters, and observational angle. Many pair studies use the dynamical friction timescale for \Tobs{} \citep{Lin2004, Bell2006, Patton2008, Masjedi08}. Another way to determine \Tobs{} is using hydrodynamical simulations of galaxy mergers and directly measuring how long close pairs or morphological signatures are observable \citep[e.g.,][]{Patton2000, Conselice2006, Kitzbichler2008, Lotz2008b, Lotz2010a, Lotz2010b, Lotz2011}. For close pairs separated by $5-20\,\kpchone$, \citet{Lotz2011} find $\langle \Tobs \rangle \approx 0.33$~\Gyr, and that $\langle \Tobs \rangle$ is essentially independent of galaxy gas fraction, and, therefore, redshift. Therefore, \Tobs{} only affects the normalization of the merger rate, not the slope as a function of redshift. 
\rr{Below, we use the merger observation timescale computed by \citet{Lotz2011}, $\langle \Tobs \rangle = 0.33\ \Gyr$. This is roughly $2\times$ longer than the \emph{minimum} expected \Tobs, namely the orbital timescale. For mergers with masses near the Milky Way mass, separated by $8\ \kpc$, the orbital timescale is $\sim0.2\ Gyr$. Nonetheless, the value of \Tobs{} introduces significant uncertainty in the normalization of the merger rate. Using $\langle \Tobs \rangle = 0.33$, the computed values for \Rmerge{} are reported in the last column in Table \ref{tabMR}.}
\middfig{\figmergerate}

Based on the corrected fraction of late-stage mergers, we find \rr{$\Rmerge \propto (1+z)^{3.8\pm0.9}$}, consistent with, albeit slightly steeper than, results of other studies that find significant evolution in the merger rate with redshift \citep[][but see \citealp{Bundy2009, Lotz2011}]{deRavel2009, Lin2008, Robaina2010}. 
\rr{The merger rate is also consistent with the expected merger rate for dark matter halos, which grow as $(1+z)^{3-4}$ \citep[e.g.,][but see \citealp{Berrier2006, Guo2008}]{Fakhouri2010}. This suggests that, at late times, massive galaxy growth traces halo growth.} 
Figure \ref{fig:mergerate} shows the fractional merger rate for late-stage mergers compared to other merger rate studies, including studies using close pairs \citep{deRavel2009, Bundy2009} and galaxy asymmetry \citep{LopezSanjuan2009, Conselice2009}. The values plotted are from \citet{Lotz2011} and take into account differences in \Tobs{} for different merger-finding methods. 

 Because the measured merger rate evolution depends on galaxy selection \citep{Lotz2011}, we limit the comparison to other mass-selected samples. All four studies in Figure \ref{fig:mergerate} use mass-selected samples, with mass limits near $\log M_*/ \Msun \gtrsim 10$, lower than our mass limit. Additionally, for the pair studies, the mass refers to the mass of the individual merging galaxies, {\emph not} the final merged galaxy, as is the case in our study. Therefore, care must be taken when comparing these results. Nonetheless, the agreement between the different methods suggests that above $\log M_*/\Msun \sim 10$, the merger rate does not depend strongly on galaxy mass. 
\rr{However, the errors on \Rmerge{} are statistical errors only and do not include uncertainties in \Tobs{}. A smaller value for \Tobs{} will increase the measured \Rmerge, and lessen the agreement between our study and previous results. }

\subsection{Mergers as a function of color and stellar mass}
\label{ssec:mcolor}
Because our sample of late-stage mergers is relatively large, we can explore the evolution of the merger rate as a function of color and stellar mass. We divide the parent sample into two mass bins, choosing the median stellar mass, $\logMsun =10.9$ as the division. \rr{Following \citet{Ilbert2013}, we further divide the sample into quiescent and star-forming galaxies, based on the rest frame near-UV (NUV)$-r^+$ and $r^+-J$ colors, where NUV corresponds to the GALEX filter at $0.23\,\mu\mathrm{m}$, and $r^+$ refers to the Subaru $r-$band. Colors are computed from the best-fit SED templates in \citet{Ilbert2013}. 
The color cuts for quiescent galaxies are: $(\mathrm{NUV} - r^+) > 3(r^+-J) + 1$ and $(\mathrm{NUV} - r^+) > 3.1$ \citep{Ilbert2013, Ilbert2010}. The color-color diagram in Figure \ref{fig:nuvrj} shows these cuts applied to our sample.}
Our analysis cannot distinguish the colors of the member galaxies in a late-stage merger. Nonetheless, we classify mergers as `wet' (gas-rich) mergers if the total merging system is star-forming, as `dry' mergers if the merger is quiescent.
\middfig{\fignuvrj}

Table \ref{tabMR} contains the corrected pair fractions within each star-forming/stellar mass bin. The fractions and merger rates reported are those within the limited star-forming/stellar mass parent sample. \rr{Note that we use the same $\Tobs$ for all sub-samples of galaxies, even though \Tobs will likely to depend on the galaxy masses and color. Within a single sub-population, however, \Tobs{} does not depend strongly on redshift, and, therefore, does not affect the evolution of the merger rate, only the normalization. The pair fractions reported in Table \ref{tabMR} are corrected for incompleteness and contamination. In addition to a dependence on redshift, \Cmerge{} is different for quiescent and star-forming galaxies because the completeness of our merger selection depends on galaxy morphology, and, therefore, color. For quiescent galaxies, we use the completeness fraction for early-type galaxies, while for star-forming galaxies, we assume the completeness of late-type mergers. For quiescent galaxies, \Cmerge{} is independent of redshift, while, for star-forming galaxies, \Cmerge{} increases by a factor of $\sim 2$  from $z\approx0.2$ to $z\approx1.0$ (see \S\ref{ssec:nocorrect}).}

Figure \ref{fig:mergecolormass} shows the corrected pair fractions for each subsample of galaxies as a function of redshift. 
From these figures it is evident that the merger rates for blue and red galaxies are significantly different. For blue galaxies $\Rmerge \sim (1+z)^{4.5\pm 1.3}$, while for red galaxies $\Rmerge$ is consistent with no evolution, $\Rmerge \sim (1+z)^{1.1\pm1.2}$. This suggests that the evolution in the merger rate for \emph{all} galaxies is driven by the evolution in the merger rate for blue galaxies and the increasing contribution of blue galaxies to the galaxy population at high redshift.
\middfig{\figmergecolormass} 
\middfig{\figmergecolorfrac}

\rr{For massive ($\logMsun > 10.9$), quiescent galaxies, $\Rmerge \sim (1+z)^{0.0\pm1.4}$, consistent with no evolution. The lack of evolution in the merger rate for massive, red galaxies agrees with results from pair studies \citep{deRavel2009, deRavel2011, Lin2008, Bundy2009, LopezSanjuan2011, LopezSanjuan2012} and simulations \citep{Kitzbichler2008}. As in \citet{Lin2008}, \citet{deRavel2009} and \citet{Darg2010b}, we find that most mergers are star-forming, and that the fraction of star-forming mergers increases significantly with redshift. Figure \ref{fig:mergecolorfrac} shows that dry, quiescent mergers mergers make up $\sim 20\%$ of all mergers in our sample, once the merger fractions are corrected for incompleteness and contamination.

It is interesting to note that Figure \ref{fig:mergecolormass} shows a weak increase in the  fractional merger rate for low mass, quiescent galaxies ($\Rmerge \propto (1+z)^{3.0\pm2.2}$). However, low-mass quiescent galaxies suffer the most from incompleteness in this $I-$band selected sample. If the average stellar mass for low mass quiescent galaxies increases as a function of redshift, then it is not surprising that the fractional merger rate in the highest redshift bin matches that of the high-mass quiescent galaxies. This suggests the evolution the low mass quiescent galaxy merger rate may be a selection effect.


The fractional merger rate for star-forming galaxies grows significantly with increasing redshift, $\Rmerge\propto (1+z)^{3.5\pm1.8}$ for high mass galaxies, and $\Rmerge\propto (1+z)^{5.4\pm1.7}$ for low mass galaxies. This demonstrates the increasing importance of wet major mergers at high redshift, in agreement with previous studies \citep{Li2008, Bundy2009, LopezSanjuan2012, LopezSanjuan2011}.  
The strongly increasing fraction of star-forming mergers also agrees with the increase in bright infrared sources as a function of redshift \citep{LeFloch2005}. This is expected as these sources are often associated with mergers \citep{Sanders1996, Kartaltepe2010}.

As with quiescent galaxies, the merger rate evolution for low mass galaxies is steeper than for high mass galaxies. Although this may be driven by incompleteness, the steepness of the merger rate for lower mass galaxies suggests a dependence of the merger rate on halo mass. Note, however, that the range of masses under study is small ($10.6 \leq \logMsun \lesssim 11$). The differences in \Rmerge{} for samples of star-forming, quiescent, low mass and high mass galaxies help explain the differences in merger rates found using different parent samples, especially the differences in \Rmerge{} between  mass-limited and luminosity-limited samples \citep[e.g.,][]{Lotz2011, Lin2004}. 


\subsection{Without incompleteness/contamination corrections}
\label{ssec:nocorrect}
\middfig{\figmergecolorfracnc}
The results in Figures \ref{fig:mergerate} through \ref{fig:mergecolorfrac} use late-stage merger fractions corrected for incompleteness and contamination. While these correction factors are based on simulations well-matched to the observed data, the corrections are still uncertain. Therefore, we perform the same analysis, excluding the completeness and contamination corrections, setting $\Cmerge=0.7$ for the line-of-sight pairs correction. This only serves to demonstrate the effect of our correction factors. The simulations clearly demonstrate the need for incompleteness and contamination corrections, and the measured merger rates are certainly invalid without any corrections. Without the incompleteness corrections, the overall late-stage merger fraction drops by a factor of $\sim 2$. Furthermore, the measured evolution in the merger rate disappears if we do not include an evolving correction for incompleteness. Figure \ref{fig:mergecolorfracnc} demonstrates that the fraction of late-stage mergers only evolves slightly with redshift, $\fecp{} \propto (1+z)^{0.8\pm0.8}$ This is in contrast to Figure \ref{fig:mergecolorfrac}, which shows a large increase in the \emph{corrected} merger fraction with redshift. 

Comparing Figures \ref{fig:mergecolorfrac} and \ref{fig:mergecolorfracnc} also shows that the fraction of quiescent mergers is unchanged by the correction factor for contamination and incompleteness. This is unsurprising since our merger sample is complete for early-type galaxies out to $z\sim1$. The completeness correction mainly affects star-forming galaxies, particularly at high redshift. Without corrections for contamination and incompleteness, the merger rates for quiescent and star-forming galaxies grow as $\Rmerge\propto (1+z)^{0.7\pm1.2}$ and $\Rmerge\propto (1+z)^{1.5\pm1.2}$, respectively. While the merger rate evolution for quiescent galaxies is unchanged, the merger rate for star-forming galaxies is significantly lower, and only marginally inconsistent with a flat merger rate. These results suggest that the non-evolving merger rate for quiescent galaxies is robust. For star-forming galaxies, the uncorrected merger rates represent a lower limit. Even if the exact values of the incompleteness we measure using mock merger images are inaccurate, our sample of late-stage mergers is demonstrably incomplete at high redshift, particularly for late-type, star-forming galaxies. Therefore, the merger rate for star-forming galaxies will evolve at least as quickly as $(1+z)^{1.5\pm1.2}$.

In the above analysis, we use a contamination correction that is independent of redshift and galaxy type (quiescent or star-forming). However, there are several contamination effects which are likely larger for star-forming galaxies than for quiescent galaxies. These contaminants may artificially boost the late-stage merger fraction at high redshift. Correcting for these effects will lower the merger rate evolution rate.  
The star-forming galaxy merger rate is particular sensitive to morphological k-corrections and the overall increase in the fraction of clumpy, star-forming galaxies at high redshift. This increase in the contamination will lead to an artificial increase in the merger rate for star-forming galaxies at high redshift. However, as demonstrated in \S\ref{ssec:contaim} and Appendix \ref{app:sims}, the overall contamination rate from non-merging galaxies is $\sim 10\%$, while the completeness is only $\sim 20\%$, and the former only depends weakly on redshift. Therefore, we expect the effects of incompleteness to dominate over any effects from contamination, and the results presented above to be robust, despite the large correction for incompleteness and contamination.
}

\section{Properties of late-stage mergers}
\label{sec:props}
To study the internal properties of late-stage mergers, we limit our parent sample to galaxies from the spectroscopic zCOSMOS survey \citep{Lilly2007, Lilly2009}. The `bright' zCOSMOS sample contains spectra for $\sim 20,000$ galaxies ($I < 22.5$). 
The zCOSMOS sample should be a subsample of the $K$-band selected sample from \citet{Ilbert2013} described above, \rr{but differences in masking mean that some zCOSMOS galaxies are missing from the photo-z sample.} To ensure no galaxies are missing, we rerun our merger selection algorithm on postage stamps generated from the zCOSMOS parent sample. The final sample is selected in the same way as that used in \citet{Kampczyk2013} and \citet{Silverman2011} to study the SFRs and AGN properties of kinematic pairs. Below, we compare the properties of late-stage mergers both to the parent zCOSMOS sample and to more widely separated kinematic pairs from the same parent sample.

Because these galaxies are observed in zCOSMOS, we use the spectroscopic redshifts, and stellar masses from \citet{Pozzetti2010} and \citet{Bolzonella2010}. \ed{These stellar masses use the spectroscopic redshifts and are in good agreement with those measured in \citet{Ilbert2013} using photometric redshifts.} 
For the analysis below, we examine galaxies with $M_* > 2.5\times 10^{10}\Msun$ in the redshift range $0.25\leq z < 1.05$, the same mass and redshift range used in kinematic pair studies in zCOSMOS \citep{Silverman2011, Kampczyk2013}. 
\rr{For the analysis of the SFR (\S\ref{ssec:sfr}), we remove all sources with a {\it Chandra} or {\it XMM} detection. This leaves a sample of $4586$ galaxies of which $154$ are classified as late-stage mergers.} We use the non-merging galaxies from the parent zCOSMOS sample as a control sample. However, we first check that the control sample is well-matched to the merger sample in stellar mass and redshift. Since AGN activity, SFR and, in this sample of galaxies, sSFR \citep[see][]{Maier2009}, are strong functions of $M_*$ and $z$, it is important that the late-stage merger sample is not biased relative to the control sample. Kolmogorov-Smirnov (K-S) tests show that both the mass and redshift distributions for late-stage mergers and the control sample are indistinguishable. 

\subsection{Star formation in late-stage mergers}
\label{ssec:sfr}
Simulations show that galaxy mergers lead to enhanced star formation \citep[e.g.,][]{Hernquist1989, DiMatteo2007, Mihos1996, Barnes1991, Springel2005, Hopkins2006}. There is also much observational evidence to support this conclusion. Galaxies identified as mergers are often bluer \citep[e.g.,][]{Kampczyk2007, Darg2010} and show enhanced UV and IR SFR \citep[e.g.,][]{Jogee2009, Robaina2009}. Spectroscopy confirms that galaxies in close pairs have higher star formation rates than isolated galaxies \citep[e.g.,][]{Barton2000, Lambas2003, Kampczyk2013, Ellison2013}. Studies of far-infrared selected galaxies show that galaxies with very high star formation rates are much more likely to be disturbed morphologically than galaxies with typical star formation rates \citep[e.g.,][]{Sanders1996, Kartaltepe2010}. In this section, we compare the SFR in late-stage mergers to that of isolated galaxies. \ed{Our control sample of isolated galaxies is simply the set of zCOSMOS galaxies that are not identified as late-stage merger candidates. }

\middfig{\figmassSFR}
Figure \ref{fig:massSFR} shows the narrow 4000-Angstrom break \citep{Balogh1999} for late stage mergers and the control sample. As expected, at fixed stellar mass, the late-stage mergers have a lower median $D_n(4000)$ than the control sample. This is indicative of recent, within the last Gyr, star formation in the late stage mergers. A K-S test shows that the distributions of $D_n(4000)$ for the late-stage mergers and the control sample are distinct. Indeed, at masses below $\logMsun=10.7$, there are almost no quiescent late-stage mergers. This indicates that a significant fraction of star formation may be associated with mergers.

\rr{We use the SFR computed from the $24\um$ flux as measured by {\it Spitzer/MIPS} \citep{Sanders2007, LeFloch2009}. The total infrared luminosity, $L_{\mathrm{IR}}$ is computed using the spectral energy distribution models from \citet{Dale2002} and the photometric redshifts from \citet{Ilbert2009} and \cite{Salvato2009} for X-ray sources. Below $z\sim 1$, the $24\um$ flux is an accurate measure of the total infrared luminosity \citep[e.g.,][]{Elbaz2010}. From the infrared luminosity, the total SFR is given by \citet{Kennicutt98}:
\begin{equation}
\label{eq:sfr}
\mathrm{SFR\ [\Msun\ yr^{-1}]} = 4.5\times 10^{-44} L_{\mathrm{IR}}/L_\odot\, .
\end{equation}
We only utilize sources with a $24\um$-detection and a measured SFR. This limits our sample to $2318$ galaxies, of which $111$ are late-stage mergers. Among galaxies with non-zero $24\um$-based SFRs, the fraction of late-stage mergers is $4.8\pm0.5\%$; in the full zCOSMOS sample, the fraction of late-stage mergers is $3.3\pm0.3\%$, suggesting an enhancement in the average SFR in late-stage mergers.
}

\middfig{\figsSFR}
\rr{
Figure \ref{fig:cumsSFR} shows the cumulative distributions of the specific star formation rate (sSFR) for our sample of mergers and galaxies in the control sample. The sSFR is calculated from the $24\um$-based SFR and the stellar mass from \citet{Pozzetti2010}. The measured stellar mass is increased by a factor $1.8$ to account for the difference in initial mass function (IMF). \citet{Pozzetti2010} use a Chabrier IMF while the SFR is computed assuming a Salpeter IMF \citep{Kennicutt98}. Figure \ref{fig:cumsSFR} shows that the distribution of sSFR is skewed to higher values for mergers compared to other galaxies in the parent sample. A K-S test shows the two distributions of sSFR are distinct. The median sSFR in late-stage mergers is enhanced by a factor of $2.1\pm 0.6$ over the sSFR in the parent sample. This demonstrates that star formation in late-stage mergers proceeds at only a moderately higher rate than star formation in isolated galaxies. Using the slightly larger and deeper photo-z parent sample with the same stellar mass and redshift limits, we find a similar enhancement in the $24\um$-derived sSFR in late-stage mergers. 

As with the merger rate, the association of late-stage mergers with star forming galaxies may be affected by contamination from clumpy, star-forming galaxies, particularly at high redshift. Our peak-finding method may identify clumpy, star-forming galaxies as late-stage mergers, which would enhance the typical sSFR in late-stage mergers. However, we show in \S\ref{ssec:contaim} that the contamination from non-merging, but clumpy, galaxies in $\sim10\%$. Furthermore, the enhancement measured here is comparable or less than that measured using other merger selection methods, strengthening our claim that the contamination from star forming, isolated galaxies is small.}

\citet{Kampczyk2013} draw similar conclusions about the [O \textsc{ii}] $\lambda3727$-derived sSFR in kinematically-selected close pairs (see their Figure 13). They find the sSFR in pairs with separations smaller than $30\,\kpchone$ is enhanced by factors of $2-4$ compared to a stellar mass- and redshift-matched control sample. \rr{Using the [O~\textsc{ii}]~$\lambda3727$ emission line as a SFR indicator \citep{Moustakas2006, Maier2009}, we also find an enhancement of $2.0\pm0.5$ in the median sSFR in late-stage mergers. The agreement between the $24\um$- and [O~\textsc{ii}]-derived sSFRs suggests that the extinction in late-stage mergers is not significantly different from that in field galaxies. However the mean sSFR computed using [O~\textsc{ii}] emission is a factor of $\sim4$ lower than the $24\um$-derived sSFR. }


We can add our sample of late-stage mergers to the more widely separated pairs in \citet{Kampczyk2013} to obtain the fraction of star formation between $0.25 < z < 1.05$ due to merging galaxies separated by less than $30\,\kpchone$. \citet{Kampczyk2013} report that $6\pm1\%$ of galaxies are in kinematic pairs \ed{with projected separations less $30\,\kpchone$. The sSFR in these galaxies is enhanced by a factor of $1.9\pm0.6$.} In the same sample, we find $3.3\pm0.3\%$ of galaxies are late-stage mergers, which have sSFRs  $2.1\pm0.6$ times above the median sSFR in the whole sample. Therefore, $18\pm5\%$ of star formation is associated with mergers, but only $8\pm5\%$ of all star formation can be considered ``excess'' star formation \emph{triggered} by mergers. This modest enhancement in the star formation due to major mergers is in agreement with other studies of visually classified mergers and close pairs \citep{Robaina2009, Jogee2009}. In addition, these results also agree with semi-analytic models \citep{Somerville2008}, which report that only $7\%$ of star formation is directly associated with major mergers. 


Since the enhancement in sSFR for late-stage mergers is small, it is important to note that these systems are not starburst galaxies. 
The small shift in the SFR for late-stage mergers is in agreement analysis by  \citet{Sargent2012}. They suggest that the sSFRs for starburst galaxies and main sequence star-forming galaxies form a double Gaussian, in which the means are offset by only a factor of $\sim 4$. This is in contrast to other definitions of starburst galaxies, requiring SFRs an order of magnitude larger than predicted by the star-forming main sequence. While the majority of starburst galaxies are major mergers \citep[e.g.,][]{Sanders1996, Kartaltepe2010, Wu1998, Cui2001}, the majority of late-stage mergers in our study are not starburst galaxies. 

\subsection{AGN fraction}
\label{ssec:agn}
\rr{In addition to triggering star formation, major mergers may drive black hole growth through AGN activity. Mergers can induce disk instabilities in coalescing galaxies that drive gas to the center of galaxies. This gas is used up both in star formation and in growing the black hole \citep[e.g.,][]{Mihos1996}. Many simulations show that the periods of most intense star formation and black hole growth occur in late-stage mergers near coalescence \citep{Mihos1996, Springel2005, DiMatteo2007, Johansson2009}. 
While simulations suggest that luminous accretion (i.e. QSOs) is dominated by major mergers, more than half of low-luminosity ($\log L_{\mathrm{bol.}}/L_\odot \lesssim 11$) AGN activity can be fueled by stochastic, non-major merger processes \citep[e.g.,][]{Hopkins2006a,Hopkins2013}. 
This is in agreement with studies of AGN host galaxies that demonstrate most AGN activity occurs in galaxies with undisturbed (non-merging) morphologies \citep[e.g.,][]{Cisternas2011, Kauffmann2004, Liu2012, Kocevski2012}. Nonetheless, there are many observations that show some enhancement in nuclear activity in galaxy pairs \citep[but see \citealp{Darg2010, Li2008, Barton2000}]{Alonso2007, Silverman2011, Ellison2011, Koss2012, Ellison2013, Woods2007}. Using spectroscopic pairs out to $z\sim 1$, \citet{Silverman2011} show the fraction of X-ray selected AGN with $10^{42} < L_{0.5-10 \mathrm{keV}} < 10^{44}$~erg~s$^{-1}$ increases from \rr{$3.8^{+0.3}_{-0.4}\%$ for isolated galaxies to $9.7^{+2.3}_{-1.7}\%$} for galaxies in pairs with a maximum projected separation of $75\ \kpc$ \ed{and a line-of-sight velocity separation less than $500\ \kmps$}. Despite this enhancement, only $\sim 25\%$ of AGN activity occurs in galaxy pairs, and an even smaller fraction, $\sim 18\%$ of AGN activity is {\it triggered} by close interactions. }

\middfig{\figAGNex}
\middfig{\figAGNfrac}

We can improve the estimate of the fraction of AGNs due to merging by including late-stage mergers with the kinematic galaxy pairs in \citet{Silverman2011}. In order to simplify the comparison, we use a parent sample identical to that of \citet{Silverman2011}, and described above ($\log M_*/\Msun > 10.4$ and $0.25 \leq z < 1.05$). We select AGN based on their X-ray flux as measured by {\it Chandra} \citep{Elvis2009}. Due to the limited area of the {\it Chandra} survey, the zCOSMOS parent sample used here only contains $3474$ galaxies of which $92$ are late-stage mergers. The X-ray sources are matched to optical/IR sources as described in \citet{Civano2012}. As in \citet{Silverman2011}, we only consider X-ray sources with total fluxes ($0.5-10$ keV) greater than $1\times 10^{-15}$~erg~cm$^{-2}\,$s$^{-1}$ and luminosities larger than $2\times 10^{42}$~erg~s$^{-1}$. The latter requirement eliminates galaxies in which the contribution from star formation to the X-ray flux is not negligible. Ninety-five percent of the X-ray sources have luminosities $L_{0.5-10 \mathrm{keV}} < 10^{44}\ \mathrm{erg~s^{-1}}$. This ensures that most of the AGN hosts we examine are not AGN-dominated in the optical/NIR, and that the derived properties, especially stellar masses are reliable \citep{Bongiorno2012, Salvato2011}. The final sample contains 164 {\it Chandra} X-ray sources, which are certain to be AGN-dominated. Figure \ref{fig:agnex} shows the $6$ late-stage mergers which are also X-ray selected AGN. 

The left panel of Figure \ref{fig:agnfrac} shows the AGN fraction as a function of pair separation for our sample of late-stage mergers(filled square) and for more widely separated kinematic pairs \citep{Silverman2011} drawn from the same parent sample. We define the AGN fraction as the fraction of late-stage mergers (close pairs) with an associated X-ray source. We compute the AGN fraction as in \citet{Silverman2011}, equation 1. This formula down-weights compulsory zCOSMOS targets, which are X-ray selected, and likely to be AGN \citep{Lilly2007}. It also accounts for the spatially varying {\it Chandra} sensitivity by weighting each AGN by the fraction of galaxies in which the measured X-ray flux is below the sensitivity, i.e., the fraction of galaxies which \emph{could} host each AGN. The values for the kinematic pairs in the left panel of Figure \ref{fig:agnfrac} are taken from the Bayesian likelihood analysis in \citet{Silverman2011}. This method takes into account contamination of the control sample with galaxies in kinematic pairs in which only one member is observed spectroscopically. Since late-stage mergers fall into a single slit, this more sophisticated approach for calculating the AGN fraction is unnecessary. We find $6$ late-stage mergers that are also X-ray selected AGN. Although the statistics are poor, the AGN fraction among late-stage mergers is $6.4\pm 2.5\%$. \rr{This is marginally consistent with the AGN fraction in the field, $3.8^{+0.3}_{-0.4}\%$ \citep{Silverman2011}. At $95\%$ confidence, we find that the AGN fraction in late-stage mergers is enhanced by less than a factor of $3.0$, with a mean value of $1.7\pm0.7$, in agreement with (although less stringent than) \citet{Cisternas2011}. 

Given the upper limit on the enhancement of AGN activity associated with late-stage mergers, we can compute an upper limit for the AGN activity triggered by mergers. Following the same procedure as \S\ref{ssec:sfr}, the late-stage merger fraction in this sample is $3.0\pm0.3\%$ and the enhancement in the AGN fraction is at most a factor of $3$ above the control sample. Therefore, the fraction of AGN activity associated with late stage-mergers is $<9.0\pm0.9\%$ and the fraction of AGN activity \emph{triggered} by late-stage mergers is at most $6.0\pm0.9\%$. Using the measured mean value for the AGN enhancement ($1.7\pm0.7$), the fraction of AGN activity triggered by late-stage mergers is 
$2\pm2\%$. 
While \citet{Silverman2011} find approximately $1/4$ of AGN are associated with kinematic pairs closer than $143\ \kpc$, only $17.8^{+8.4}_{-7.4}\%$ of AGN can be directly contributed to the pair interaction. Combining the kinematic pairs with our late-stage mergers gives a total fraction of AGN activity triggered by mergers of $\sim 20\pm8\%$. As expected, including late-stage mergers does not significantly increase the fraction of AGN activity contributed to major mergers.

The right panel in Figure \ref{fig:agnfrac} shows the AGN fraction in pairs divided into two redshift bins, $0.25\geq z <0.65$ and $0.65\geq z<1.05$. Among late-stage mergers, all $6$ X-ray selected AGN occur above $z=0.65$. This boosts the AGN fraction at high redshift to $11.5\pm4.2\%$, which is statistically above the value for the field, and comparable to the boost seen for kinematic pairs separated by less than $75\ \kpc$ \citep{Silverman2011}. Note that in the right panel of Figure \ref{fig:agnfrac}, the field AGN fraction is not corrected for contamination from kinematic pairs in which only one galaxy is observed. Since kinematic pairs do show an enhanced AGN fraction, correcting for this contamination will likely lower the field AGN fraction by $\sim 0.5-1\%$. Below $z\sim0.65$, none of our late-stage mergers are also X-ray AGN. The error bar in Figure \ref{fig:agnfrac} shows the $1\sigma$ upper limit for the AGN fraction, which is consistent with the AGN fraction in the field, albeit with large uncertainty. Although our results rule out a decrease in the merger rate at close separations, and suggest some enhancement in the AGN fraction at $z\gtrsim 0.7$, a larger sample is required to determine if any enhancement in the AGN rate for late-stage mergers is statistically significant.
 }


%

From the X-ray and optical images alone, it is unclear if any of the late-stage mergers with AGN are dual AGNs. The possible dual AGN, CID-42 \citep{Comerford2009, Civano2010}, is excluded from the sample above since the measured stellar mass of the system is below $2.5\times 10^{10}\Msun$, however our method does select CID-42 as a late-stage merger. We have examined the spectra for the $6$ mergers with AGN and find no evidence for velocity offsets, suggesting that only one of the black holes in the system is actively accreting. Nonetheless, since the late-stage mergers are selected to have two, concentrated central cores, this sample would be well-suited to searches for galaxies with dual AGN.

\subsubsection{AGN in the photo-z sample}
\label{sssec:agnphoto}
\rr{
We can check the fraction of mergers with AGN using the photo-z sample. For the photo-z sample, we use the same stellar mass and redshift cuts as the spectroscopic sample. However, we exclude X-ray sources that are best-fit by a Type I AGN/QSO template \citep{Salvato2011}.
Since we identify late-stage mergers based on the flux ratio of two central peaks, our method is poorly-suited to selecting companions of bright Type I AGN, in which the optical flux is dominated by a single point source. This means that the AGN fraction reported here is for lower luminosity, obscured AGN, not bright Type I AGN. 

Using the photo-z sample, we find $8$ late-stage mergers that are {\it Chandra}-detected X-ray AGN with $L_{[0.5-10\ \mathrm{keV}]} > 2\times 10^{42}$~erg~s$^{-1}$, $4$ of which are also identified in the spectroscopic sample. The missing $2$ galaxies are excluded because of differences in the masking in the $K-$band and $I-$band selected catalogs, used for the photo-z and spec-z samples \citep[see][]{Ilbert2013,Lilly2007}. These $8$ late-stage mergers yield an AGN fraction of $5\pm2\%$, consistent with the AGN fraction fraction found above and that in the field. 
Because of the high degree of overlap between the photo-z and spec-z galaxy samples, all our results on the AGN fraction are highly correlated and do not add significant statistical power. Complete coverage of the COSMOS area with {\it Chandra} will improve the statistics of these results by a factor $\sim 2$ and yield a larger sample of AGN associated with mergers (Civano et al., {\it in prep}).}

As with the star formation rates, we find that, although AGN activity may be slightly enhanced in late-stage mergers, mergers do not drive the majority of AGN activity, and, hence, black hole growth. 
Including late-stage mergers along with more widely separated pairs, only $\sim 20\%$ of AGN activity is triggered by mergers, and late-stage mergers are responsible for at most $6\%$ of AGN activity. 
The small increase in AGN activity associated with close pairs is in agreement with previous studies \citep{Ellison2013, Alonso2007} and suggests minor mergers or secular processes within galaxies drive the majority of low-luminosity AGN activity. 
Similarly, while the star formation rate in late-stage mergers is typically enhanced compared to a control sample, the enhancement is less than a factor of $2$ and 
only $8\pm5\%$ of star formation can be contributed to kinematic pairs or late-stage mergers. 
The similarity between the AGN enhancement and SFR enhancement in merging galaxies \citep[see also][]{Silverman2011} suggests that star formation and AGN activity are physically coupled, as is expected from simulations of gas-rich major mergers \citep{Mihos1996, Barnes1992, Hernquist1989, Hopkins2006, Hopkins2008a}.

It is possible that our focus on late-stage mergers ignores other phases of galaxy merging with larger enhancements in AGN activity and star formation \citep[see][]{Scudder2012}. The bright QSO phase may occur when the two nuclei are closer to coalescence \citep[e.g.][]{DiMatteo2005, Hopkins2008a}. However, by combining kinematic pairs with late-stage mergers, we can observe galaxy mergers from separations of $100\,\kpchone$ until shortly before coalescence. \rr{Furthermore, the $0.5-8$ keV X-ray selection for AGN may exclude highly obscured AGN, which would be visible in the infrared \citep[e.g.,][]{Satyapal2014} or harder X-ray bands \citep[e.g.,][]{Koss2011}. Using WISE-selected AGN in SDSS, \citet{Satyapal2014} find an enhancement of the AGN fraction in nearby post-merger galaxies of a factor $10-20$, suggesting the AGN in late-stage mergers are highly obscured.}  Nonetheless, the low enhancement in X-ray selected AGN and star formation activity in late-stage mergers and kinematic pairs reconfirms the results that galaxy and black hole growth are not solely driven by major mergers.

\section{Summary}
\label{sec:discussion}
Although mergers of dark matter halos underpin theories of structure and galaxy formation, the actual role of galaxy mergers is less clear. In this work, we seek to expand the study of merging galaxies to galaxy pairs with small separations. By including late-stage mergers with samples of more widely separated (as yet not merging) pairs, we can obtain a better understanding of the role of mergers in galaxy evolution since $z\approx1$.
To that end, we develop a method to identify late-stage galaxy mergers using {\it HST} images. 

We utilize a high-pass filter which easily detects the bright, concentrated, central cores of both member galaxies of a merger before coalescence. \rr{By implementing limits on the flux ratio and brightness of the measured peaks, we are able to produce a clean sample of $2055$ galaxy mergers from COSMOS ACS $I-$band images of galaxies brighter than $I=23$. These late-stage mergers have two intact galaxy nuclei that are separated by less than $8$~\kpc. If we restrict the parent sample to a mass-complete ($\log M_*/M_\odot > 10.6$) sample of galaxies in the redshift range $0.25<z<1.0$, with pair separations between $2.2$ and $8$ kpc, we find $136$ late-stage mergers, which represents $2.3\pm 0.2\%$ of the massive galaxy population, or $4.8\pm0.5\%$ when corrected for contamination and incompleteness. The sample of late-stage mergers identified here is distinct from other samples of merging galaxies, such as kinematic pairs, and morphologically disrupted galaxies identified by CAS or Gini/M20. 

We create mock images of mergers by placing two real galaxies in one postage stamp and use these to test the completeness and contamination in our sample. Although the sample suffers little from contamination ($10\%$ from clumpy, non-merging galaxies and $20\%$ from minor mergers), we only successfully select $\sim20\%$ of all major mergers, and the selection efficiency decreases with increasing redshift. Our method is most successful for mergers between concentrated early-type galaxies, selecting $80\%$ of all simulated mergers, independent of redshift. }

Using the sample of late-stage mergers we study both the evolution of the merger rate and the properties of merging galaxies. Our results can be summarized as follows:
\begin{itemize}
\item For galaxies with stellar masses above $\logMsun > 10.6$, we find that the fraction of mergers evolves as $\fecp \propto (1+z)^{3.8\pm0.9}$ when corrected for incompleteness, and contamination from minor mergers, non-mergers and line-of-sight superpositions. Despite uncertainties in the sample completeness, and the merger timescale, \Tobs{}, the normalization of the fractional merger rate, \Rmerge{}, agrees well with that found in previous studies. The measured evolution in the merger rate becomes significantly flatter if we remove the redshift-dependent correction for incompleteness of the sample, $\Rmerge \propto (1+z)^{0.8\pm0.8}$.
\item \rr{Dividing the sample into quiescent, star-forming, low mass, and high mass galaxies, we find that the merger rate for star-forming galaxies is a strong function of redshift, $\Rmerge \propto (1+z)^{4.5\pm1.3}$, while that for quiescent galaxies is a mild function of redshift, consistent with no evolution, $(1+z)^{1.1\pm1.2}$.  Therefore, among massive galaxies, the increase in the total merger rate is driven by the increase in the merger rate for star-forming galaxies and by the increasing fraction of massive star-forming galaxies at high redshift. Lower mass ($10.6 < \logMsun < 10.9$) galaxies also exhibit a steeper merger rate evolution than higher mass ($\logMsun > 10.9$) galaxies: $(1+z)^{5.1\pm1.3}$ compared to $(1+z)^{2.7\pm1.1}$. These results use different corrections for completeness for star-forming (late-type) mergers and quiescent (early-type) mergers. Although the merger rate slopes are not as steep without the corrections for incompleteness, the merger rate for star-forming (low-mass) galaxies still evolves more than that of quiescent (high-mass) galaxies. This shows that the differences in the merger rates as a function of stellar mass and SFR are robust. Furthermore, these differences suggest that measurements of the merger rate as a function of redshift are very sensitive to the sample of galaxies.
}
\item Examining the properties of late-stage mergers, we find that the SFR in late-stage mergers with $\log M_*/\Msun > 10.4$ is enhanced by a factor of \rr{$2.1\pm0.6$} compared to non-interacting galaxies. This is similar to the enhancement found for kinematic galaxy pairs using the same parent sample \citep{Kampczyk2013}. Only $18\pm5\%$ of star formation between $z=0.25$ and $z=1.05$ is associated with late-stage mergers or pairs separated by less than $30\,\kpchone$. However, the excess star formation that can be attributed to major mergers is only half of that.
\item The AGN fraction in late-stage mergers at $z>0.5$ is enhanced by a factor of $2.2\pm0.8$ compared to the field. For the entire redshift range, $0.25<z<1.05$, we do not measure a statistically significant enhancement in AGN activity. At most, the AGN activity in late-stage mergers between $0.25<z<1.05$ is enhanced by a factor of $3$ above the activity in field galaxies. Together with more widely separated pairs, $20\pm8\%$ of AGN activity is \ed{induced by} mergers at separations less than $143\,\kpc$. The fraction of AGN triggered by late-stage mergers and kinematic pairs is similar to the fraction of SFR activity triggered by the same class of mergers. This suggests that the processes responsible for star formation and AGN activity in major mergers are coupled, indicating a co-evolution scheme \citep{Jahnke2009, Cisternas2011, Schramm2013}.
\end{itemize}

The measurement of the blue galaxy merger rate is particularly sensitive to morphological k-corrections and the increasing fraction of blue, star-forming, clumpy galaxies at high redshift. Because galaxies appear clumpier at blue rest frame wavelengths and the entire galaxy population contains more clumpy, star-forming galaxies at high redshift, we expect our peak-finding method to detect more late-stage merger candidates at high redshift. We plan to address this by applying our peak-finding method near-infrared data WFC3 {\it HST} data from the CANDELS survey \citep{Koekemoer2011, Grogin2011}. \rr{Performing this study at longer wavelengths may also help increase the completeness of our merger sample. Galaxies appear more bulge-dominated and concentrated at longer wavelengths and our merger-finding method is significantly more sensitive to mergers between concentrated, early-type galaxies than mergers between late-type galaxies.}

Although we have examined the SFRs and X-ray emission of  late-stage mergers, resolved properties of the mergers require additional data. For instance, we have a sample of $\sim20$ late-stage mergers with significant X-ray detections and $2$ concentrated central cores. Although the X-ray detection cannot resolve the merging galaxies, these sources provide an excellent parent sample for spectroscopic searches for dual AGN \citep[e.g.,][]{Comerford2009, Liu2010, Civano2010}. \ed{By focusing only on X-ray AGN, we are only studying a subset of AGN. There are many AGN selected in \emph{IRAC} \citep{Donley2012}, radio \citep{Smolcic2008}, and optical/infrared \citep{Fiore2008, Dey2008}. Increasing the sample size of AGN will increase the number of AGN in late-stage mergers. Further work is needed to determine if the \emph{fraction} of AGN activity associated with late-stage mergers will also increase using AGN selected in the optical, infrared, or radio. }

Obtaining a sample of late-stage mergers between $1 \lesssim z \lesssim 2$ would allow us to continue to measure the evolution of the merger rate at higher redshifts. Recent spectroscopic studies suggest that the merger rate increases quickly beyond $z\sim1$ \citep[e.g.,][]{LopezSanjuan2013, Tasca2014}. Expanding the sample to higher redshift requires high resolution and signal-to-noise data in the near infrared. Switching to the longer wavelengths eliminates the effects of morphological $k-$corrections, as there is evidence that galaxies typically have more structure at shorter wavelengths \citep[e.g.,][]{Kuchinski2000}. Near-infrared WFC3 {\it HST} data from the CANDELS survey \citep{Grogin2011, Koekemoer2011} could be used for such a study. Additionally, since the star formation and AGN activity also continue to grow with redshift, expanding the sample of mergers to higher redshift will increase the statistical significance of the sample of late-stage mergers with ongoing star formation and AGN activity. This will help determine the role of major mergers in the growth of galaxies and super-massive black holes at their peak epoch of formation.

\acknowledgments We thank the referee for their insightful comments which greatly improved this work. We also thank Richard Massey and Kevin Bundy for helpful comments on a draft. This work was supported by World Premier International Research Center Initiative (WPI Initiative), MEXT, Japan. ST acknowledges support from the Lundbeck Foundation. The Dark Cosmology Centre is funded by the Danish National Research Foundation. This research has made use of the NASA/IPAC Infrared Science Archive, which is operated by the Jet Propulsion Laboratory, California Institute of Technology, under contract with the National Aeronautics and Space Administration. This research made use of APLpy, an open-source plotting package for Python hosted at http://aplpy.github.com

\appendix
\section{Appendix A: Simulated merger images}
\label{app:sims}
\rr{In order to test our merger-finding algorithms, we create a sample of mock mergers by coadding real galaxy images from our original sample to create new postage stamps. These fake merger images have the same properties as the real galaxy images and allow us to test both the completeness and contamination of our merger selection. In particular, the mock mergers will have the same redshifts, magnitudes, morphologies, and merger ratios  as the real galaxy population. Because the fraction of real mergers is small, they represent a small contamination in our sample of mock mergers.  While these simulations are realistic in many ways, they do not include any structural changes wrought by the merger in the images. However, since our method is sensitive to only the brightest features in merging galaxies, this omission is likely unimportant. Below, we focus on mergers in which the total mass is larger than $4\times10^{10}\Msun$ and the redshift is between $0.25 < z < 1.0$. These are the cuts in \S\ref{sec:mergerrates} and ensure the sample is complete at all redshifts.

\middfig{\figmocks}
\middfig{\figzcomplete} 
To create a mock merger postage stamps, we randomly select $2400$ galaxies from our photo-z sample. For each selected galaxy, we select at random a second galaxy at approximately the same photometric redshift ($\Delta z < 0.02$) as the first selected galaxy. This ignores any contamination from chance superpositions at widely different redshifts, which we address statistically in \S\ref{ssec:contaim}. We then coadd the {\it HST}/ACS postage stamp images of these galaxies. For each galaxy pair, we make $8$ postage stamps with different separations between the galaxy centers, spanning from $0.1$ to $10$ kpc. Using the ZEST morphology parameter \citep{Scarlata2007}, we create $3$ additional samples of $1200$ mock mergers with specific morphologies: a sample of mergers between two early-type galaxies (ZEST=1), a sample between late-type galaxies (ZEST=2), and a mixed sample (ZEST=1 and 2). We apply the median ring filter to all these mock images and examine the results. 
Figure \ref{fig:mocks} shows four example mock merger images. The circles denote the two galaxies which have been super-imposed. Because of the coaddition, these mock images are a factor of $\sqrt{2}$ nosier than the original images. However, because we are creating merger images by coadding images from our sample, the mock mergers are up to a factor of $2$ brighter than the real galaxies in the sample. These effects approximately cancel out and we neglect the differences in signal-to-noise between the mock mergers and the real galaxy images. 

Figure \ref{fig:mocks} also demonstrates that our method does not detect all mock mergers (blue $\times$s in Figure \ref{fig:mocks}). Our selection is particularly incomplete for mergers among late-type galaxies, in which neither galaxy has a dominant bulge. The left panel of Figure \ref{fig:zcomplete} indicates our completeness as a function of redshift (black solid line). In this figure, we examine the completeness for major mergers ($> 1:4$) with separations between $2.2$ and $8.0$ \kpc, without applying any cuts in flux ratio, while the right panel includes cuts in flux ratio (see below). Beyond $z\sim 0.5$, our method only detects $20\%$ of the mock late-stage mergers. Dividing the sample by the ZEST morphology of the merging galaxies shows that the completeness is a strong function of morphology. The median ring filter selects $80\%$ of early-type mergers, but only $\sim 40\%$ of late-type mergers. This is expected, since our method requires a strong central bulge in order to detect a peak. However, in comparing the merger rates of early and late type galaxies, it is important to note their very different completeness fractions. Furthermore, the decrease in the completeness up to $z\sim0.5$, suggests that our merger rate evolution may be underestimated, as we are missing $\sim 2\times$ as many galaxies at high redshift than at low redshift. 

Figure \ref{fig:scomplete} shows the completeness as a function of pair separation, similarly divided by galaxy morphology. For all morphologies, the completeness is independent of pair separation beyond $\sim 2-3\ \kpc$. This is driven by the size of the median ring filter ($2.2\ \kpc$ at $z=1$). The lower panel shows the fractional error in our measurement of the pair separation. For large separations, the separation is well-measured. However, for small real separations, the measured separation is typically too large. By cutting off the pair separations  at $2.2~\kpc$, we introduce some contamination from smaller separation pairs, particularly at lower redshifts. Determining the exact fraction of this contamination would require knowledge of the spectrum of real pair separations, which we have not included in these simulations. However, we can assume that the number of mergers at small separations will be smaller than those at large separations, so we expect this contamination to be small. In particular, the contamination from other galaxy structures and minor mergers is likely to be much larger, and can be measured using our simulated merger images.
\middfig{\figscomplete}

Because our mock merger images use real galaxies, we can ascertain how often galactic sub-structures, such as bars, spiral arms, and star-forming clumps, are selected by the median ring filter. These features are typically fainter than galaxies, and by implementing a cut on the ratio of the peak flux to the total galaxy flux, we can eliminate a substantial fraction of the contamination for galaxy substructure. Figure \ref{fig:fluxfrac} shows the distribution of peak fluxes for peaks associated with merging galaxies and peaks associated with galaxy substructure. The substructure peaks are typically fainter. In order to eliminate these detections, we require that the detected peaks contain at least $3\%$ of the total galaxy flux. This reduces the contamination by extraneous peaks to $\sim 10\%$, while keeping the completeness of \emph{detected} peaks at $\sim 80\%$. Lowering this threshold to $0.5\%$ does not significantly affect our results. Before applying our method to different imaging data, similar simulations should be conducted in order to determine the threshold value. 

\middfig{\figfluxfrac}
\middfig{\figfcontaim}

In addition to contamination from star-forming clumps and galactic substructure, our merger finding method is somewhat sensitive to minor mergers. Since we are only interested in studying late-stage major mergers, it is important to understand the contamination from minor mergers. To create a mock merger sample with a realistic fraction of minor mergers, we build a sample of 2400 mock mergers in which one member of the pair is brighter than $I=20.5$. This ensures that, from a sample of galaxies with $I<23$, we include a realistic spectrum of merger ratios down to $1:10$. Note that our algorithm only measures the flux ratio of the merger, \emph{not} the underlying mass ratio, which requires color information about the separate galaxies. However, in our mock merger images, we find that the \emph{real} flux ratio of a merger is well-correlated with the mass ratio, and that the flux ratio of $0.25$ corresponds approximately to a mass ratio of $0.25$. 

Figure \ref{fig:fcontaim} shows the contamination rates for our sample of detected mergers, including major and minor mergers, as well as false positives, i.e. mergers made of star forming clumps instead of the two galaxies inserted into the mock image. We exclude mergers with a observed flux ratio smaller than $0.25$. Overall, $70\%$ of the detected late-stage mergers are real major mergers, and only $10\%$ of mergers are false positives, as expected from the peak flux to total flux cut explained above. The total fraction of minor mergers is $20\%$, and the contamination is worse at lower flux ratios. The lower panel in Figure \ref{fig:fcontaim} shows the fractional error in the measured flux ratio. While the measured flux ratio correlates with the real flux ratio, the errors are extremely large, particularly at small real flux ratios. Better measurements of the flux ratio could be obtained by fitting a late-stage merger with two realistic galaxy profiles centered on each detected peak. However, in this work, we only use the measured flux ratio to discriminate between major and minor mergers. By only selecting mergers with a flux ratio larger than $0.25$, we only eliminate $15\%$ of the detected major mergers and $30\%$ of minor mergers. Note, that the median ring filter is less sensitive to minor mergers than major mergers, and that many minor mergers are eliminated by the peak flux to total flux cut of $3\%$. Both of these effects further help to limit contamination from minor mergers. 

Taken together, our cuts in peak to total flux ratio, and peak to peak flux ratio, do affect the overall completeness, particularly the late-type galaxy merger completeness. After implementing the flux ratio cuts, the overall completeness drops to $20-25\%$ (see the right panel of Figure \ref{fig:zcomplete}), with most of the decrease coming from late-type mergers. However, these cuts are important since they significantly decrease the contamination from star forming clumps and minor mergers. Using the results of Figure \ref{fig:zcomplete} and \ref{fig:fcontaim}, we can correct the measured late-stage merger fractions for incompleteness and contamination and use the corrected fractions to determine the merger rate evolution. In studying the internal properties of late-stage mergers, we cannot include a correction for incompleteness. However, in this case, it is more important to have a minimally contaminated sample of late-stage mergers, as significant contamination will mask any differences between the field population and the merger population. 

}

\section{Appendix B: Comparison to CAS and Gini-$M_{20}$}
\label{app:casgini}
\middfig{\fignotgini}
\middfig{\fignotasym}
\middfig{\figisgini}
\middfig{\figisasym}
\rr{
In order to better understand the poor overlap between our sample of late-stage mergers and mergers selected based on their Gini (G), $M_{20}$ and asymmetry (A) values, we examine a small random set of galaxy images. 
A galaxy is considered a merger by the Gini-$M_{20}$ method if $G > -0.12 M_{20} + 0.38$ \citep{Lotz2008}. A galaxy is considered a merger based on its asymmetry if $A > 0.35$ \citep{Conselice2003}. The morphology measurements G, $M_{20}$, and asymmetry ($A$) values are taken from \citet{Cassata2005}. Note that the deblending done by \citet{Cassata2005} leads to different values for the morphology metrics than those derived directly from the images shown here. However, because we are looking for merging, deblending close pairs may not always be desirable, and we include differences in the deblending as part of our comparison.

The Gini-$M_{20}$ and asymmetry merger selections were designed to work in the rest frame $B-$band at low redshift \citep[e.g.][]{Conselice2003, Lotz2008}. In order to compare the results to higher redshifts, morphological k-corrections need to be taken into account. By using galaxies at redshifts above $z\sim0.6$, the observed $I-$band images are close to the rest frame $B-$band images and corrections to the measured G, $M_{20}$, and $A$ can be neglected. We do include a correction of $\delta A = 0.05$ for the effect of surface brightness dimming at high redshift \citep[see][]{Conselice2009, Conselice2003, Conselice2003a}.

Figure \ref{fig:notgini} shows late-stage mergers selected by our method that are not selected by the Gini-$M_{20}$ criterion. Panels $b$, $c$, $d$, $g$, and $h$ show galaxies with Gini and $M_{20}$ values close to the division line. In panel $a$, the detected peaks are well-separated from the main galaxy and are likely a separate system. In panels $e$ and $f$, the central peaks are well enough separated to be deblended before measuring the morphology. This will lower the $M_{20}$ coefficient in particular. In general, the galaxies our merger method selects are highly concentrated, which leads to lower $M_{20}$ values than for other mergers. 

Figure \ref{fig:notasym} shows late-stage mergers with $A<0.35$ that are not considered mergers based on their asymmetry. Galaxies in panels $c$, $d$, $g$, and $h$ likely have low asymmetry values due to differences in deblending. However, it is worth noting that an equal mass merger between two similar galaxies will be symmetric about an $180^\circ$ rotation, which may contribute to the low A values in the case of galaxies in panels $d$ and $g$. As with the galaxies in Figure \ref{fig:notgini}, the galaxies shown here are highly concentrated, which also tends to lower the asymmetry value.

Figure \ref{fig:isgini} shows instead the galaxies detected as mergers by the Gini-$M_{20}$ criterion but not selected as late-stage mergers. Panels $b$, $c$, $d$, $g$, and $h$ show galaxies with only one bright central peak. The galaxy in panel $d$ may have two bright nuclei, but they are not separable by our method. The peaks detected in panels $a$ and $e$ are too faint compared to their host galaxy to be included by our method. Our method would characterize these galaxies as star-forming, not merging. The peaks detected in panel $f$ are separated by slightly more than $8\ \kpc$ and are therefore excluded from our sample. 

Figure \ref{fig:isasym} shows galaxies which are selected as mergers based on their asymmetry, but not by our median ring filter method. The asymmetric features in almost all of these galaxies are too faint to be detected by our method. The mergers in panels $e$ and $g$ both have flux ratios below our threshold of $0.25$ and would be considered minor mergers. Indeed, many of the galaxies not detected by our method but detected by other methods are minor mergers, in which one component is significantly fainter than another, or the companion is no longer visible and only other tracers of the merger remain. This suggests our merger finding method may be complementary to other merger finding methods better suited to finding minor mergers.
}

\bibliographystyle{apj}
\bibliography{lackner_short}

\tailfig{\figdemo}
\tailfig{\figexample}
\tailfig{\figreject}
\tailfig{\figginiM}
\tailfig{\figconcenasym}
\tailfig{\figmergerate}
\tailfig{\figmasscomp}
\tailfig{\figmergecolormass}
\tailfig{\figmergecolorfrac}
\tailfig{\figmassSFR}
\tailfig{\figsSFR}
\tailfig{\figAGNex}
\tailfig{\figAGNfrac}
\tailfig{\figfcontaim}
\tailfig{\figzcomplete}
\tailfig{\tablepairs}
\tailfig{\tablemr}

\end{document}